%% file: arxiv.tex
\RequirePackage{lineno}
\documentclass[prd,amsmath,amssymb]{revtex4}

\usepackage{rotating}
\usepackage{subfigure}
\usepackage{graphicx}
\usepackage{dcolumn}
\usepackage{bm}
\usepackage{psfig}
\usepackage{epsfig}
\usepackage{psfrag}
\usepackage{needspace/needspace}

\def\met{{\mbox{$E\kern-0.57em\raise0.19ex\hbox{/}_{T}$}}}
\def\DZ{D\O\ }

\def\ifb{fb$^{-1}$}

\def\pp{p\bar{p}}
\def\bb{b\bar{b}}
\def\cc{c\bar{c}}
\def\cs{c\bar{s}}
\def\ttbar{$t\bar{t}$}

\def\lvbb{$\ell\nu\bb$}
\def\llbb{$\ell\ell\bb$}
\def\vvbb{$\nu\nu\bb$}

\def\ra{\rightarrow}

\def\wlv{W\ra\ell\nu}
\def\zll{Z\ra\ell\ell}
\def\zvv{Z\ra\nu\nu}

\def\wcs{W\ra c\bar{s}}
\def\zcc{Z\ra\cc}
\def\zbb{Z\ra\bb}

\def\etal{{\it et.~al.}}

\def\gev{~Ge\kern -0.05em V\kern -0.1em /$c^2$}
\def\dgev{Ge\kern -0.05em V\kern -0.1em /$c^2$}
\newcommand{\GeV} {\ensuremath{\mathrm{Ge\kern -0.1em V}}}

\def\alpgen{{\sc alpgen}}
\def\pythia{{\sc pythia}}
\def\comphep{{\sc comphep}}
\def\mcfm{{\sc MCFM}}

\def\lumimin{7.5} 
\def\lumimax{8.4}

\def\vznlo{4.4}   
\def\vznloe{0.3}  

\def\wwnlo{11.3}  
\def\wwnloe{0.8}  
\def\wznlo{3.2}	  
\def\wznloe{0.2}  
\def\zznlo{1.2}   
\def\zznloe{0.1}  

\def\vzRF{5.0}

\def\vzRFstat{1.0}

\def\vzRFsystu{1.3}
\def\vzRFsystd{1.2}

\def\vzRFexpnsigma{2.9}
\def\vzRFnsigma{3.3}

\def\vzresult{$\sigma(VZ)=\vzRF\pm\vzRFstat$~(stat) $^{+\vzRFsystu}_{-\vzRFsystd}$~(syst)~pb}

\def\wzRF{5.9}

\def\wzRFstat{1.4}
\def\wzRFsyst{0.7}

\def\wzresult{$\sigma(WZ)$~=~\wzRF~
              $\pm$~\wzRFstat~(stat)~
              $\pm$~\wzRFsyst~(syst)~pb}

\def\zzRF{0.45}

\def\zzRFstat{0.61}
\def\zzRFsyst{1.2}

\def\zzresult{$\sigma(ZZ)$~=~\zzRF~
              $\pm$~\zzRFstat~(stat)~
              $\pm$~\zzRFsyst~(syst)~pb}

\begin{document}


\begin{figure}
\leftline{\includegraphics[scale=0.5]{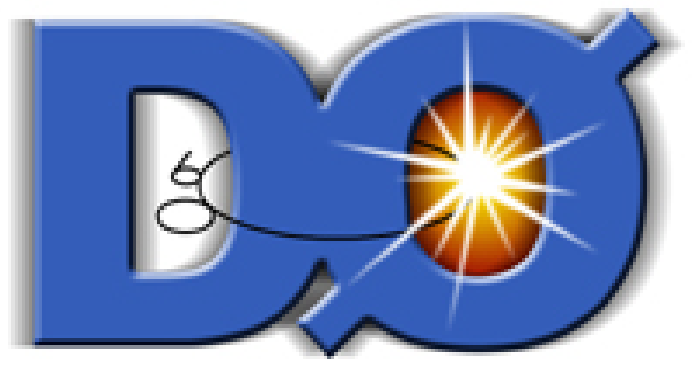}
\hfill D\O~Note 6260-CONF}
\end{figure}

\title{Evidence for $WZ$ and $ZZ$ production in final states with $b$-tagged jets \vspace*{2.0cm}}

 \author{The D\O\ Collaboration}
 \affiliation{URL http://www-d0.fnal.gov}


\date{November 15, 2011}

\begin{abstract}
  \vspace*{3.0cm} We present evidence for the combined production of
  $VZ$ ($V=W$ or $Z$) events in final states containing charged
  leptons (electrons or muons) or neutrinos, and heavy flavor jets,
  using data collected by the D\O\ detector at the Fermilab Tevatron
  Collider.  The analyzed samples correspond to \lumimin\ to \lumimax\
  \ifb~of $\pp$~collisions at $\sqrt{s}=1.96$ TeV.  Assuming the ratio
  of the production cross sections $\sigma(WZ)$ and $\sigma(ZZ)$ as
  predicted by the standard model, we measure the total $VZ$ cross
  section to be \vzresult. This corresponds to a significance of
  \vzRFnsigma~standard deviations above the background only
  hypothesis.  Furthermore, we have separately measured the cross
  sections for the $WZ$ and $ZZ$ processes to be \wzresult~and
  \zzresult, in agreement with the standard model prediction.
\vspace*{4.0cm}
\end{abstract}

\maketitle
\centerline{\em Preliminary Results for the HCP 2011 Conference}

\newpage
\section{Introduction}
\label{intro}

\def\citeall{\cite{dzWHl,dzZHv,dzZHl}}

The production of $VV$ ($V=W,Z$) boson pairs provides an important
test of the electroweak sector of the standard model (SM).  In
$p\bar{p}$ collisions at $\sqrt{s}=1.96$ TeV, the next-to-leading
order (NLO) SM cross sections for these processes are
$\sigma(WW)=\wwnlo\pm\wwnloe$~pb, $\sigma(WZ)=\wznlo\pm\wznloe$~pb and
$\sigma(ZZ)=\zznlo\pm\zznloe$~pb~\cite{dibo}.  Measuring a significant
departure in cross section or deviations in the predicted kinematic
distributions would indicate the presence of anomalous gauge boson
couplings~\cite{bib:anocoups} or new particles in extensions of the
SM~\cite{bib:newphen}.  The $VV$ production in $\pp$ collisions at the
Fermilab Tevatron Collider has been observed in fully leptonic decay
modes~\cite{bib:leptonic} and in semi-leptonic decay
modes~\cite{bib:hadronic}, where the combined $WW+WZ$ cross section
was measured.

In this note we report evidence of $WZ$ and $ZZ$ production in final
states where one of the $Z$ boson decays to $\bb$ (although there is
some signal contribution from $\wcs$, $\zcc$) and the other weak boson
decays to charged leptons or neutrinos ($\wlv$, $\zvv$, or $\zll$,
with $\ell=e,\mu$).  This analysis is also relevant as a proving
ground for the searches for a low-mass Higgs boson produced in
association with a weak boson and decaying into a $\bb$ pair
\cite{bib:higgs}, which share the exact same selection
criteria and analysis techniques.

\section{Summary of Contributing Analyses}
\label{analyses}

This result is the combination of three analyses~\citeall~outlined in
Table~\ref{tab:chans}.  These analyses utilize data corresponding to
integrated luminosities ranging from \lumimin\ to \lumimax~\ifb,
collected by the D0 detector \cite{dzero} at the Fermilab Tevatron
Collider.  They are organized into multiple sub-channels for different
configurations of final state particles.  To facilitate proper
combination of signals, the analyses were constructed to use mutually
exclusive event selections.

In the \lvbb~analysis~\cite{dzWHl}, events containing an isolated
electron or muon, and two or three jets are selected.  The presence of
a neutrino from the $W$ decay is inferred from a large imbalance of
transverse momentum ($\met$).  The \vvbb~analysis~\cite{dzZHv} selects
events containing large $\met$ and exactly two jets.  Finally, in the
\llbb~analysis \cite{dzZHl}, events are required to contain two
electrons or two muons and at least two jets.  In the \lvbb\ and
\llbb\ analyses, each lepton flavor of the $W/Z$ boson decay
($\ell=e,\mu$) is treated as an independent channel.  To ensure that
the samples for the different analyses do not overlap, the \lvbb\
analysis rejects events in which a second isolated electron or muon is
identified, and the \vvbb\ analysis rejects events in which any
isolated electrons or muons are identified.

To isolate the $\zbb$ decays, an algorithm for identifying jets
consistent with the decay of a heavy-flavor quark is applied to each
jet ($b$-tagging).  Several kinematic variables sensitive to displaced
decay vertices and jet tracks with large transverse impact parameters
relative to the hard-scatter vertices are combined in a $b$-tagging
discriminant based on boosted decision trees.  This algorithm is an
upgraded version of the neural network $b$-tagging tool used
previously~\cite{bib:btagnn}.  By adjusting the minimum requirement on
the $b$-tagging output, a spectrum of increasingly stringent
$b$-tagging operating points is achieved.
Each of the analyses is separated into two groups: a double-tag (DT)
group in which two of the jets are $b$-tagged with a loose tag
requirement (\lvbb\ and \vvbb) or one loose and one tight tag
requirement (\llbb); and an orthogonal single-tag (ST) group in which
only one jet has a loose (\lvbb\ and \vvbb) or tight (\llbb) $b$-tag.
A typical per-jet efficiency and fake rate for the loose (tight)
$b$-tag selection is about 80\% (50\%) and 10\% (0.5\%), respectively.
The corresponding efficiency for jets from $c$-quarks is 45\% (12\%).
Furthermore, the \lvbb\ and \vvbb\ analyses use the output from the
$b$-tagging alogrithm as input to final discriminants.  The signal in
the DT sample is mainly composed of events with $\zbb$ decays with
smaller contributions from $\zcc$ and $\wcs$ decays.  In the ST
sample, which places a much less stringent requirement on the $b$-jet
content of the event, the contributions from the three decay modes are
comparable.  All three analyses use multivariate discriminants (MVA)
based on decision trees as the final variables for extracting the $VZ$
signal from the backgrounds.

\begin{table}[bp]
\caption{\label{tab:chans}List of analysis channels and their corresponding
integrated luminosities.  See Sect.~\ref{analyses} for details ($\ell=e, \mu$).}
\begin{ruledtabular}
\begin{tabular}{lcc}
\\
Channel                      & Luminosity (\ifb) & Reference\\\hline
\lvbb,~ ST/DT, 2/3 jets      & 7.5               & \cite{dzWHl}\\
\vvbb,~ ST/DT  2 jets        & 8.4               & \cite{dzZHv}\\
\llbb,~  ST/DT $\geq$ 2 jets & 7.5               & \cite{dzZHl}\\
\end{tabular}
\end{ruledtabular}
\end{table}

The backgrounds from multijet production are measured from control
samples in the data. The other backgrounds are generated by
\alpgen~\cite{alpgen} and \comphep~\cite{comphep}, with
\pythia~\cite{pythia} providing parton-showering and hadronization.
The primary background is from $W/Z$+jets, and is modeled with
\alpgen. The \lvbb\ and \llbb\ analyses normalize these background to
the data, whereas the \vvbb\ analysis normalizes them to the
prediction from \alpgen.  The fraction of the $W/Z$+jets in which the
jets arise from heavy quarks ($b$ or $c$) is obtained from NLO
calculations using \mcfm~\cite{mcfm}.  The background from \ttbar\
events is normalized to the approximate NNLO cross section
\cite{ttbar_xsec}.  The $s$-channel and $t$-channel cross sections for
the production of single-top quarks are from
approximate NNLO+NNLL calculations \cite{schan_top_xsec} and 
approximate NNNLO+NLL calculations \cite{tchan_top_xsec}, 
respectively.
The background from $WW$ events
is normalized to NLO calculations from \mcfm~\cite{dibo}.

\section{Systematic Uncertainties}

The main sources of systematic uncertainty varies between the
different analyses~\citeall.  Here we summarize only the largest
contributions.  The \vvbb\ and \lvbb\ analyses carry an uncertainty on
the integrated luminosity of 6.1\%~\cite{lumi}, while the overall
normalization of the \llbb\ analysis is determined from the NNLO
$Z/\gamma^*$ cross section \cite{dyxsec} in data events near the peak
of $\zll$ decays.  The uncertainty from the identification and
measurement of jets is $\sim$7\%.  The uncertainty arising from the
$b$-tagging rate ranges from 1 to 10\%.  All analyses include
uncertainties associated with lepton measurement and acceptances,
which range from 1 to 9\% depending on the final state.  The largest
contribution for all analyses is the theoretical uncertainty on the
background cross sections at 7-20\% depending on the analysis channel
and specific background.  The uncertainty on the expected multijet
background is dominated by the statistics of the data sample from
which it is estimated.  In addition, the analyses incorporate
shape-dependent uncertainties on the kinematics of the dominant
backgrounds.  These shapes are derived from the potential variations
of the MVA distributions due to generator and background modeling
uncertainties.  Further details on the systematic uncertainties are
given in
Tables~\ref{tab:d0systwh}-\ref{tab:d0llbb1}.  All systematic
uncertainties originating from a common source are held to be
correlated, as detailed in Table \ref{tab:corr}.

\section{Measurement of the $WZ+ZZ$ Cross Section}

The total $VZ$ cross section is determined from a fit of the MVA
distributions of the background and signal samples to the data.  The
ratio of the $WZ$ and $ZZ$ cross sections is fixed to its SM
prediction.  The production of $WW$ events is considered as a
background.  This fit is performed simultaneously on the distributions
in all sub-channels by minimizing a negative log likelihood ratio
function with respect to the signal cross section and variations in
the systematic uncertainties ~\cite{bib:poisson}. This function is
constructed from terms for Poisson fluctuations in the data, and a
Gaussian prior for each systematic uncertainty.  The magnitude of the
systematic uncertainties is effectively constrained by the regions of
the MVA distribution with low signal over background ratio.  Different
uncertainties are assumed to be mutually independent, but those common
to multiple sub-channels are assumed to be 100\% correlated.

The combined fit for the total $VZ$ cross section distributions yields
\vzresult.  This measurement is consistent with the NLO SM prediction
of $\sigma(VZ)=\vznlo\pm\vznloe$~pb~\cite{dibo}.  
To visualize the sensitivity of the combined analysis, we calculate
the signal over background (s/b) in each bin of the MVA distributions
from the contributing analyses.  Bins with similar s/b are then
combined to produce a single distribution, shown in Figure~\ref{fig:rfsub}.
In Figure~\ref{fig:mjj} we show the distributions of the invariant
mass of the dijet system after adjusting the signal and background
predictions according to the results of the fit.  Figure
\ref{fig:mjj_sub} shows the background subtracted dijet mass
distributions after the fit.  Distributions of the MVA and dijet
mass for the contributing analyses can be found in the
Appendix.

\begin{figure*}[tbp]
\begin{centering}
\includegraphics[height=0.2\textheight]{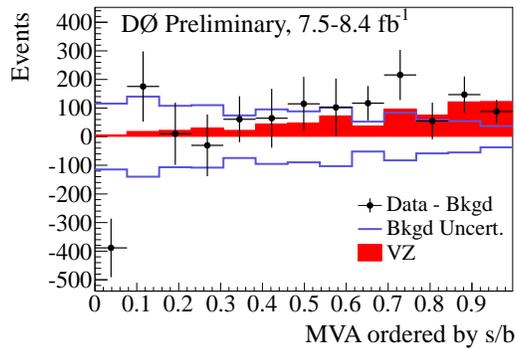}
\end{centering} 
\caption{\label{fig:rfsub} Comparison of the measured $VZ$ signal
  (filled histogram) to background-subtracted data (points).  The
  background has been fit to the data in the hypothesis that both
  signal and background are present. Also shown is the $\pm$1 standard
  deviation uncertainty on the fitted background.}
\end{figure*}

We estimate the statistical significance of the measured $VZ$ signal
by performing the same measurement on an ensemble of
pseudo-experiments drawn from the background only hypothesis.  Figure
\ref{fig:xsec_pe} shows the distribution of cross sections obtained
from the background only pseudo-experiments compared to the cross
section measured from data.  The significance is found to be
\vzRFnsigma~standard deviatons (s.d.).  The expected significance is
\vzRFexpnsigma.  Also shown in Figure~\ref{fig:xsec_pe} is the
distribution of cross sections obtained from pseudo-experiments drawn
from the SM signal+background hypothesis.  It is also interesting to
compare the distributions of the negative log-likelihood ratio (LLR)
test statistic~\cite{cls} for the two hypotheses to the LLR observed
in data.  We display the results of this comparison in Figure
\ref{fig:llr}.

\begin{figure*}[tp]
\begin{centering}
\begin{tabular}{ccc}
\includegraphics[width=2.4in]{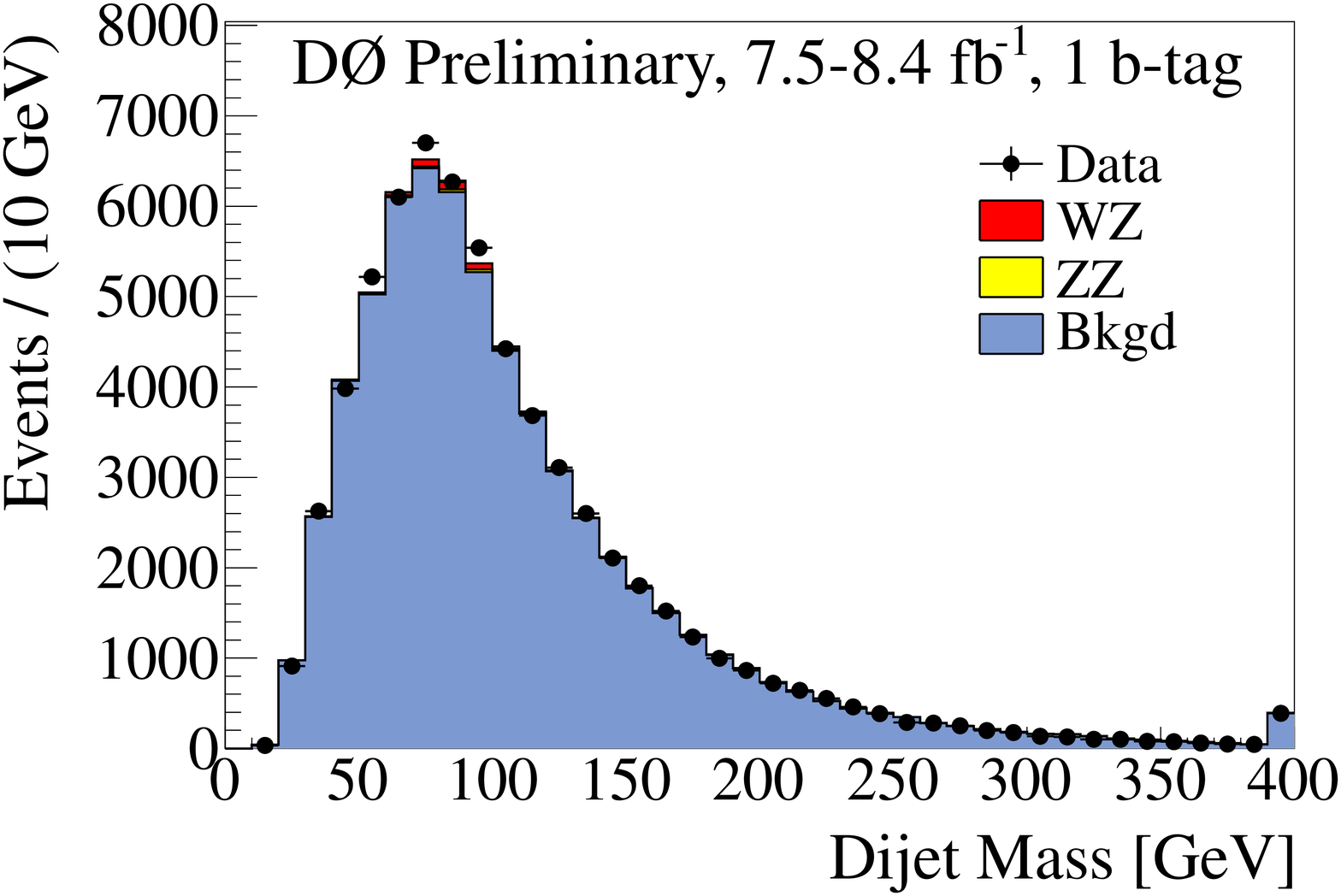}  &
\includegraphics[width=2.4in]{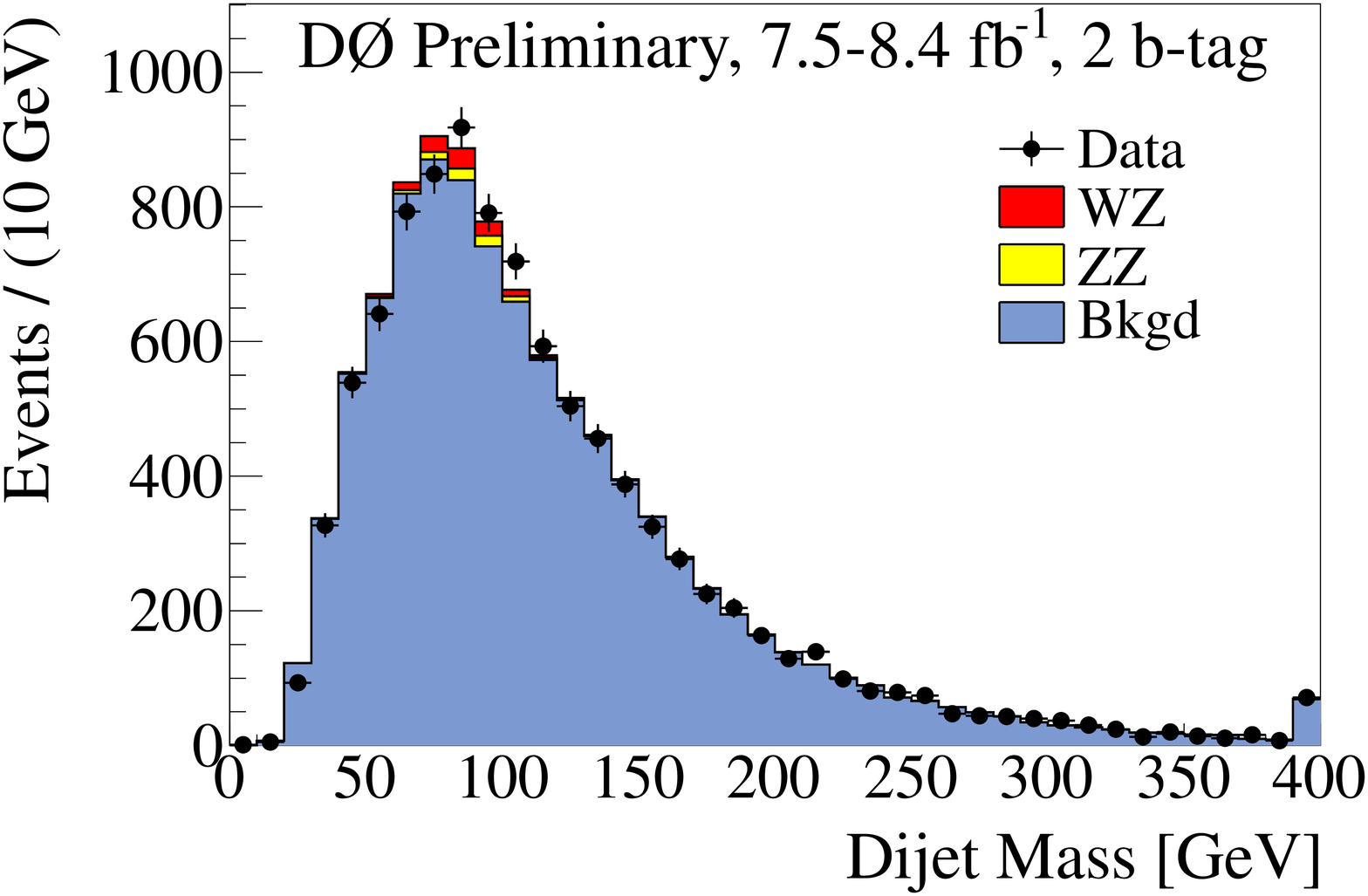} & 
\includegraphics[width=2.4in]{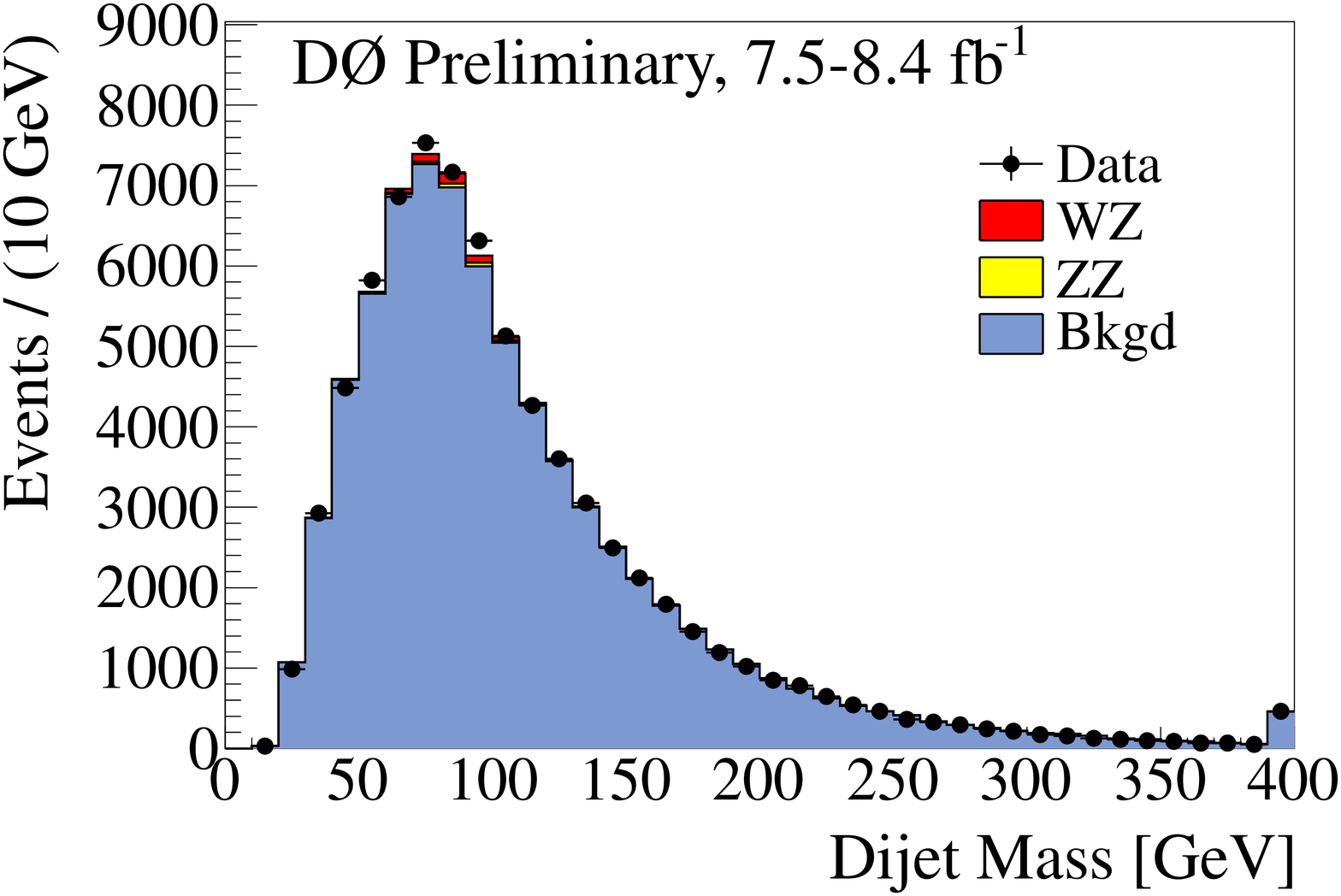} \\
{\bf (a)} & {\bf (b)}  & {\bf (c)}
\end{tabular} 
\end{centering} 
\caption{\label{fig:mjj} Comparison of the fitted signal+background to
	data in the dijet mass distribution (summed over all channels) 
        for the (a) ST, and (b) DT sub-channels; and (c) the sum of the
        ST and DT sub-channels.  Events with a dijet mass greater
        than 400 GeV are included in the last bin of the distribution.}
\end{figure*}

\begin{figure*}[tp]
\begin{centering}
\begin{tabular}{ccc}
\includegraphics[width=2.4in]{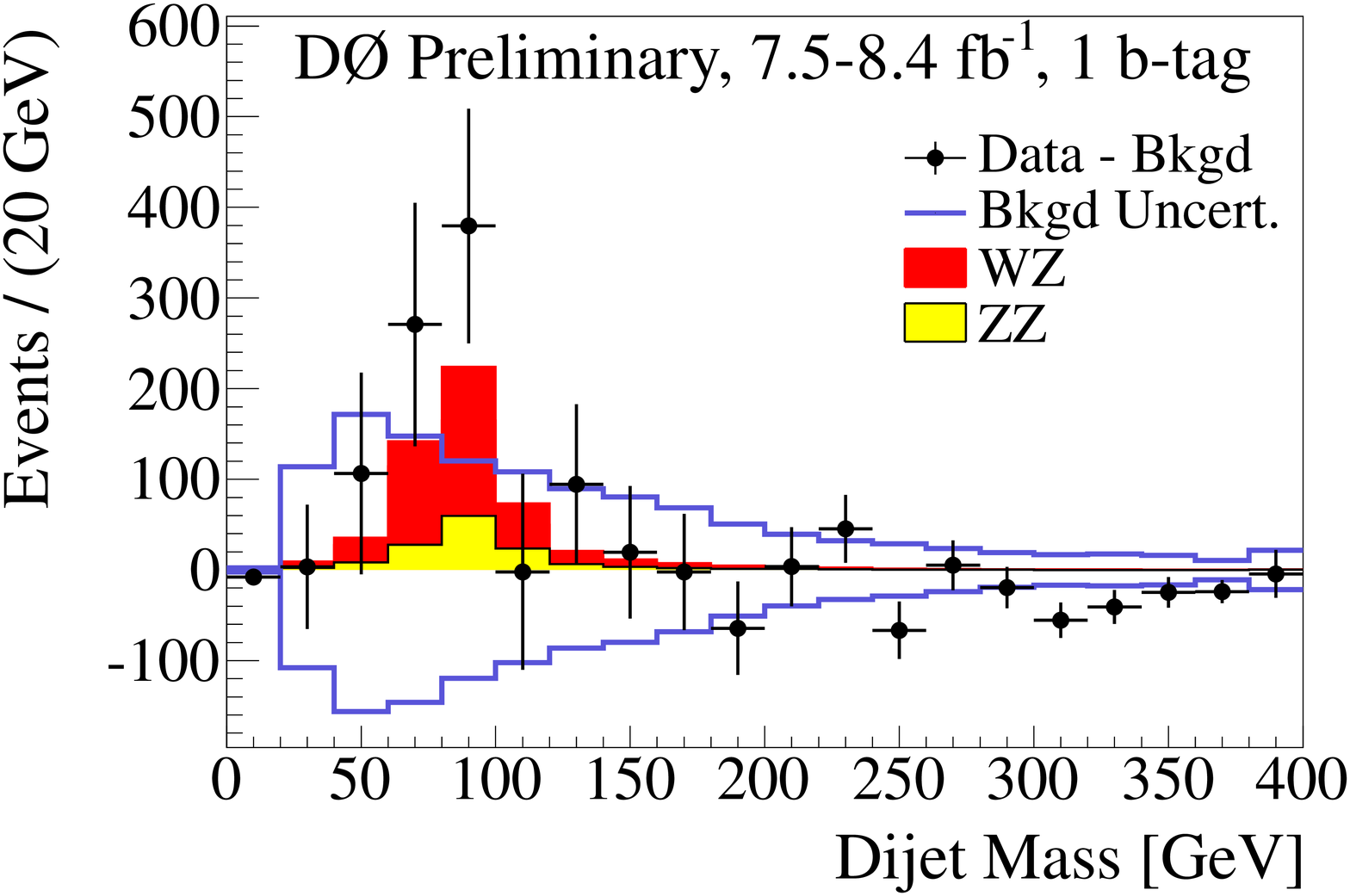}  &
\includegraphics[width=2.4in]{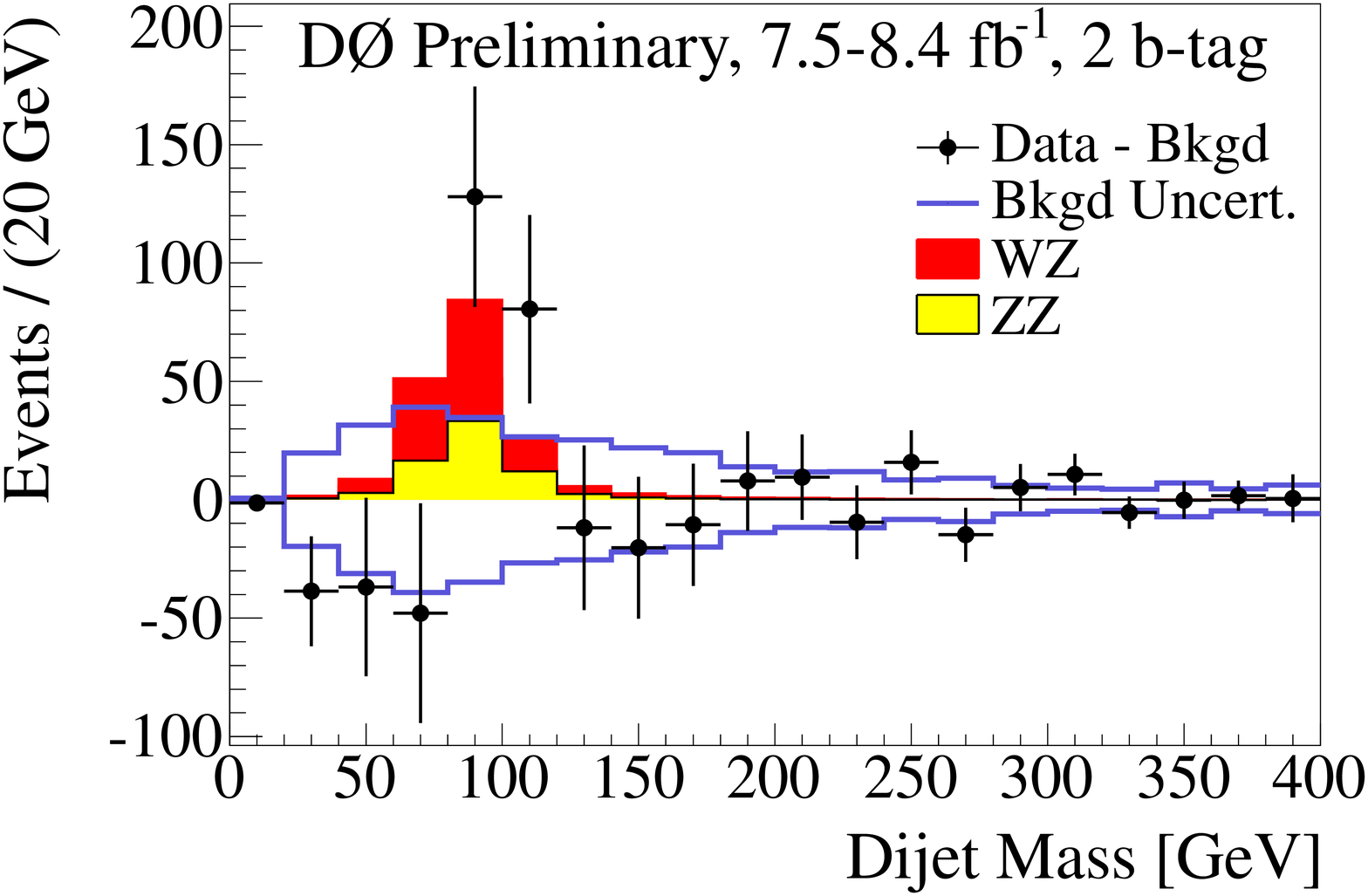} & 
\includegraphics[width=2.4in]{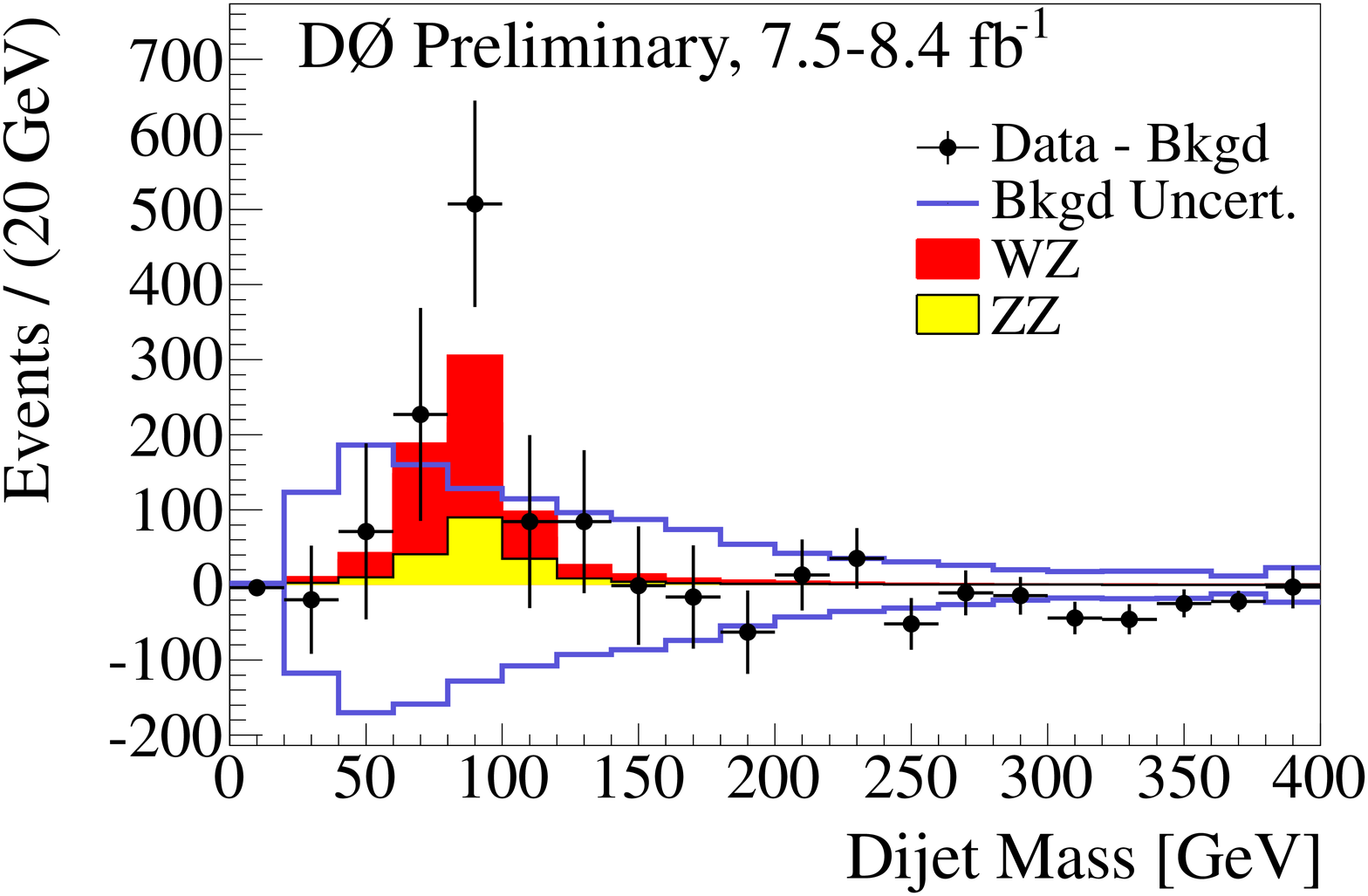} \\
{\bf (a)} & {\bf (b)}  & {\bf (c)}
\end{tabular} 
\end{centering} 
\caption{\label{fig:mjj_sub} Comparison of the measured $WZ$ and
	$ZZ$ signals (filled histograms) to background-subtracted data
(points) in the dijet mass distribution (summed over all channels) 
        for the (a) ST, and (b) DT sub-channels; and (c) the sum of the
        ST and DT sub-channels. Also shown is the $\pm$1 standard deviation
        uncertainty on the fitted background.  Events with a dijet mass greater
        than 400 GeV are included in the last bin of the distribution.}
\end{figure*}

\begin{figure*}[tp]
\begin{tabular}{cc}
\includegraphics[height=0.21\textheight]{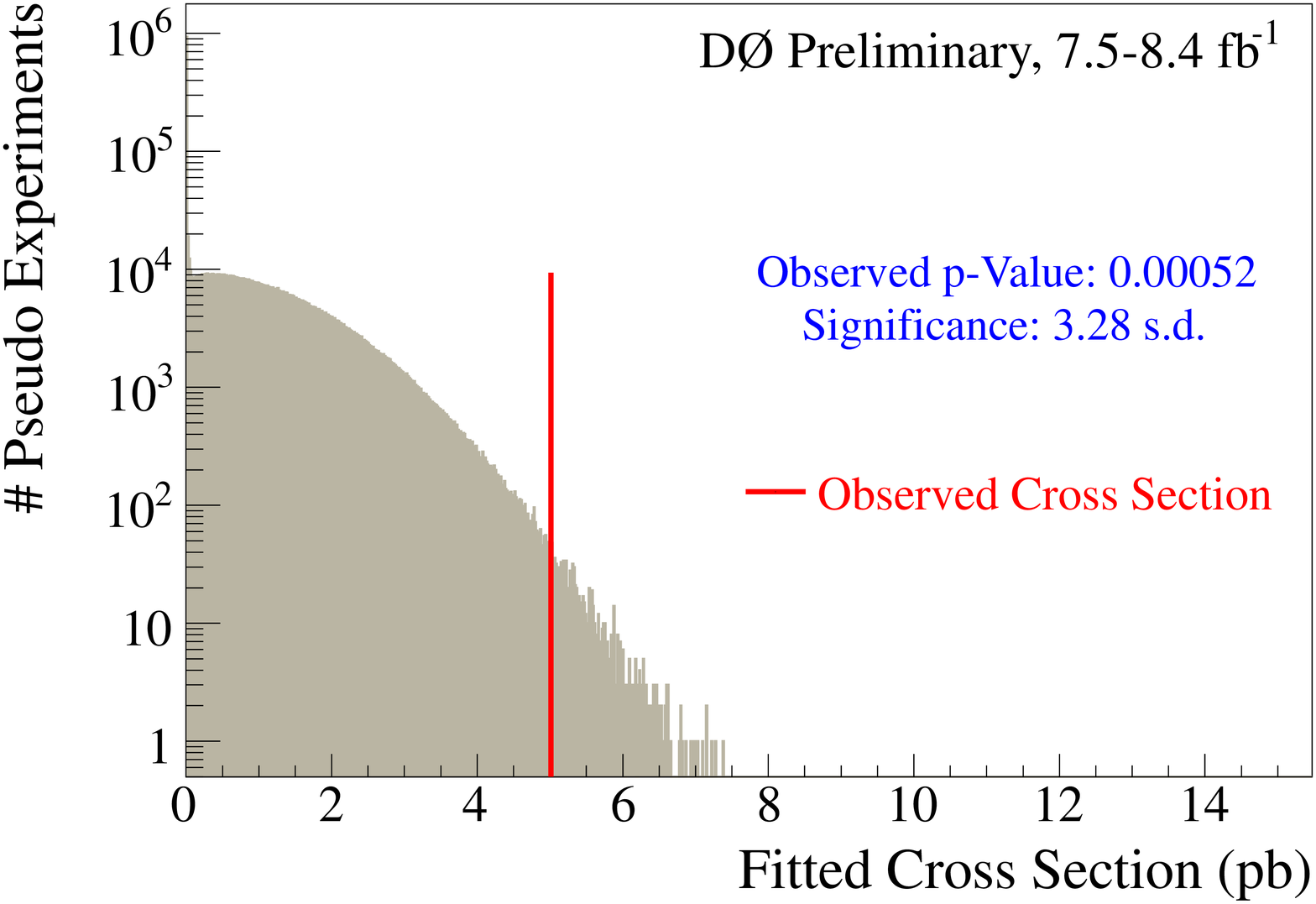}  &
\includegraphics[height=0.21\textheight]{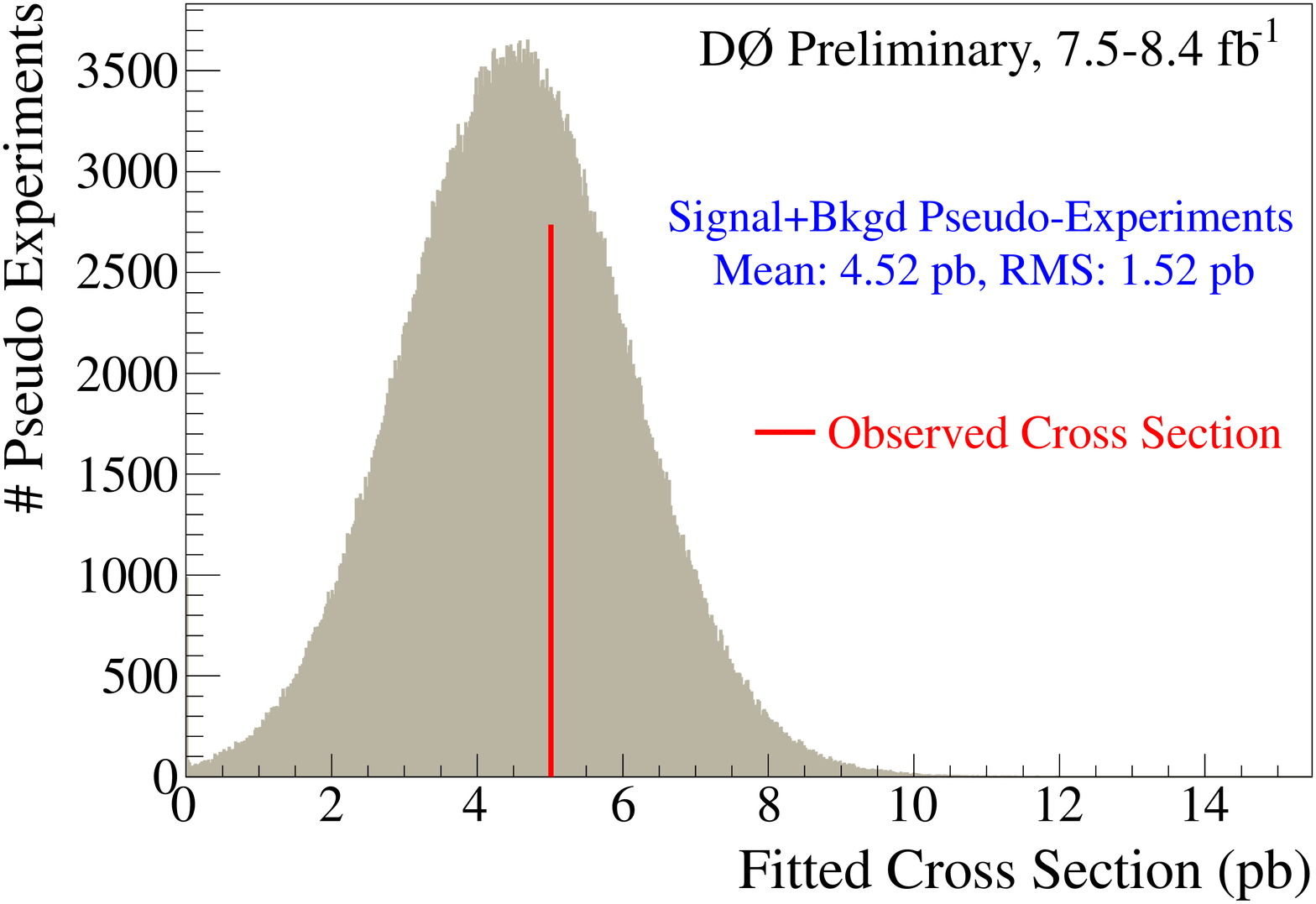} \\
{\bf (a)} & {\bf (b)}
\end{tabular} 
\caption{\label{fig:xsec_pe} Distribution of $VZ$ cross sections obtained
from 
(a) background-only pseudo-experiments and
(b) signal+background pseudo-experiments.
The observed cross section from the data (vertical red line)
is also shown.}
\end{figure*}

\begin{figure*}[htbp]
\includegraphics[height=0.21\textheight]{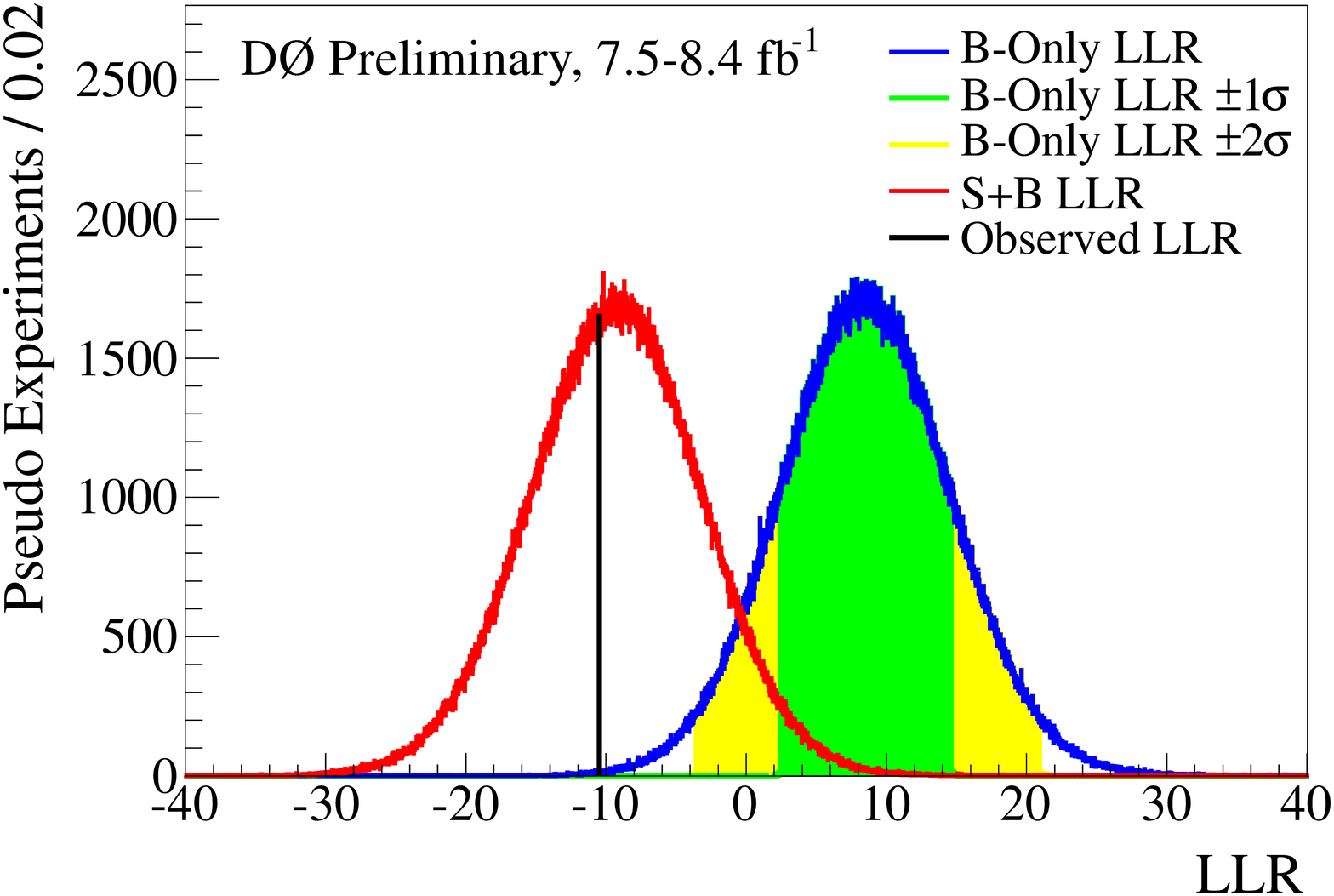}
  \caption{\label{fig:llr} LLR distributions obtained from
    background-only and signal+background pseudo-experiments compared
    to the LLR obtained from the data.}
\end{figure*}

We also perform the fit with the signal divided into its separate
$WZ$ and $ZZ$ components, which are allowed to float independently.
The result of this simultaneous fit of $\sigma(WZ)$ and $\sigma(ZZ)$
using the MVA output distributions is shown in Figure.~\ref{fig:fit_xsec_2d}.
It yields \wzresult~and~\zzresult.  These results are to be compared
to the NLO predictions of  
$\sigma(WZ)=$~\wznlo~$\pm$~\wznloe~pb and
$\sigma(ZZ)=$~\zznlo~$\pm$~\zznloe~pb.

\begin{figure}[tbp] 
\begin{centering}
\includegraphics[width=3.4in]{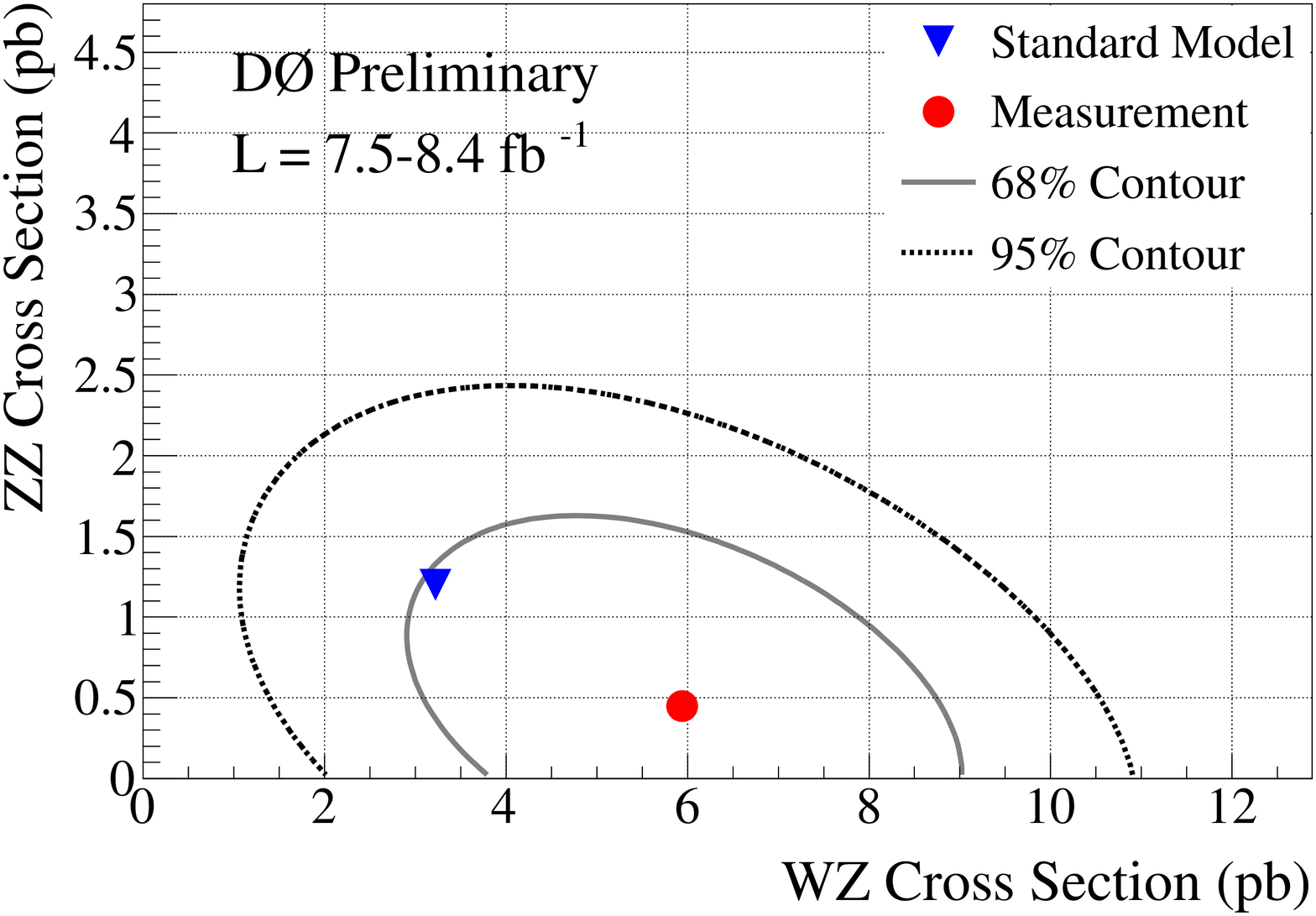}
\caption{Results from the simultaneous fit of $\sigma(WZ)$ and
      $\sigma(ZZ)$.  The plot shows
      the best fit value with 68\% and 95\% uncertainty ellipses and
      the NLO SM prediction. }
\label{fig:fit_xsec_2d}
\end{centering}
\end{figure}

\needspace{8\baselineskip}
\section{Summary}

In summary, we have combined analyses in the \lvbb, \vvbb, and \llbb\
($\ell=e,~\mu$) final states to obtain evidence with a significance of
\vzRFnsigma~s.d., for the production of $VZ$ ($V=W$ or $Z$) events.
The analyzed samples correspond to \lumimin\ to \lumimax\ \ifb~of
$\pp$~collisions at $\sqrt{s}=1.96$ TeV.
We measure the total cross section for $VZ$ production to be
\vzresult.  Furthermore, we have separately measured the cross
sections for the $WZ$ and $ZZ$ processes to be \wzresult~and
\zzresult, in agreement with the SM predictions.  These results
demonstrate the ability of the D0 experiment to measure a signal
containing two heavy-flavor jets in a background-dominated final state
directly relevant to low mass Higgs searches.

\newpage
\begin{acknowledgments}
\input{acknowledgement}
\end{acknowledgments}

\newpage
\appendix

\section{Additional Material}

\input{lvbb-sys}

\input{vvbb-sys}

\input{llbb-sys}

\begin{table}[htpb]
  \caption{\label{tab:corr}The correlation matrix for the analysis
    channels.  Uncertainties marked with an $\times$ are considered
    100\% correlated across the affected channels.  Otherwise
    the uncertainties are not considered correlated, or do not
    apply to the specific channel.  The systematic uncertainties
     on the background cross section ($\sigma$) and the normalization
    are each subdivided according to the different background
    processes in each analysis. }
\begin{ruledtabular}
\begin{tabular}{lcccc}\\
Source                     & \lvbb\      & \vvbb\        & \llbb\      \\ \hline
Luminosity                 & $\times$   & $\times$   &              \\
Normalization	           &            &	     &              \\
Jet Energy Scale           & $\times$   & $\times$   & $\times$     \\
Jet ID                     & $\times$   & $\times$   & $\times$     \\
Electron ID/Trigger        & $\times$   & $\times$   & $\times$     \\
Muon ID/Trigger            & $\times$   & $\times$   & $\times$     \\
$b$-Jet Tagging            & $\times$   & $\times$   & $\times$     \\
Background $\sigma$        & $\times$   & $\times$   & $\times$     \\
Background Modeling        &            &            &              \\
Multijet Background        &            &            &              \\
Signal $\sigma$            & $\times$   & $\times$   &  $\times$    \\
\hline
\\
\end{tabular}
\end{ruledtabular}
\end{table}

\newpage

\begin{figure*}[tbp]
\begin{centering}
\begin{tabular}{ccc}
\includegraphics[width=2.4in]{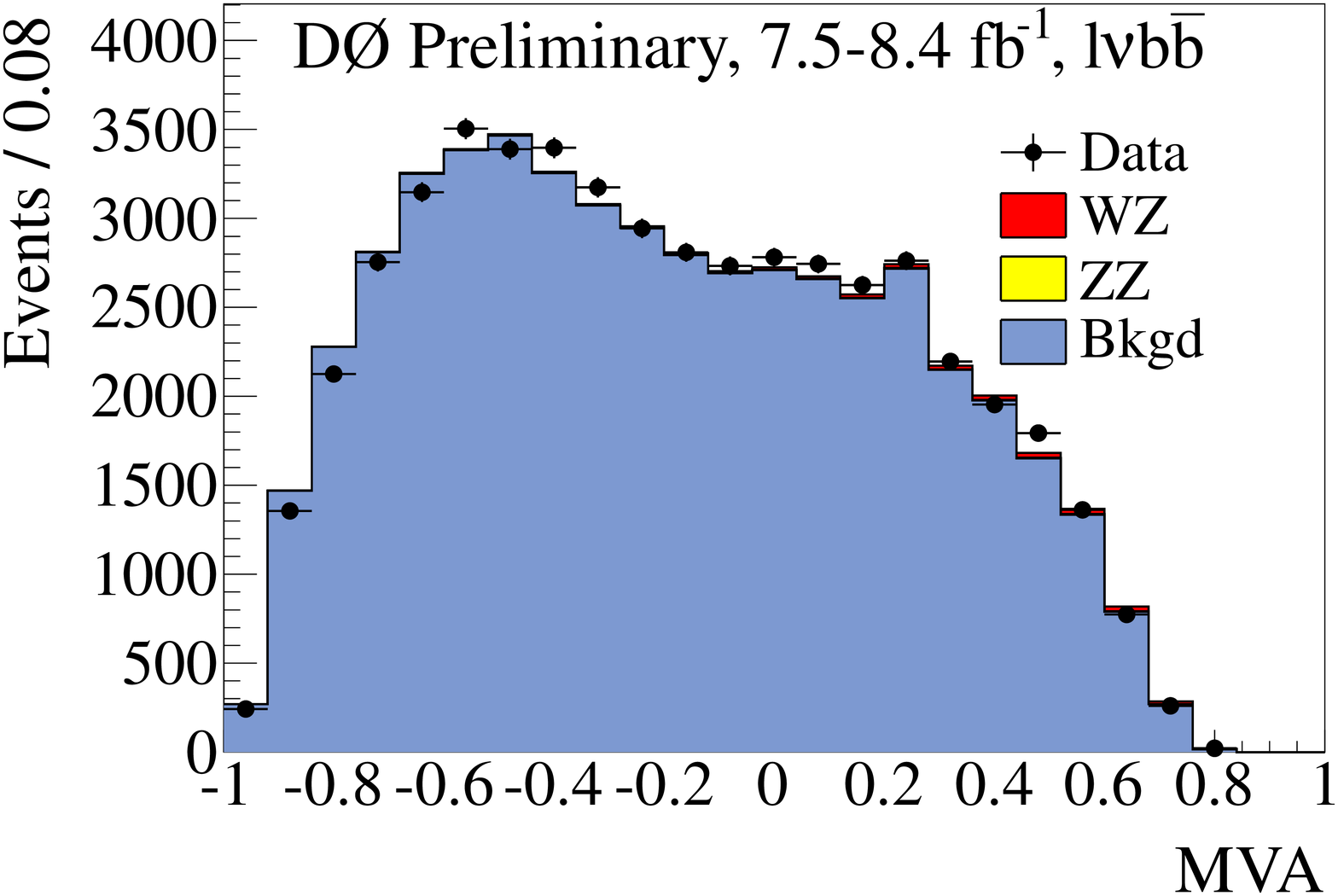}  &
\includegraphics[width=2.4in]{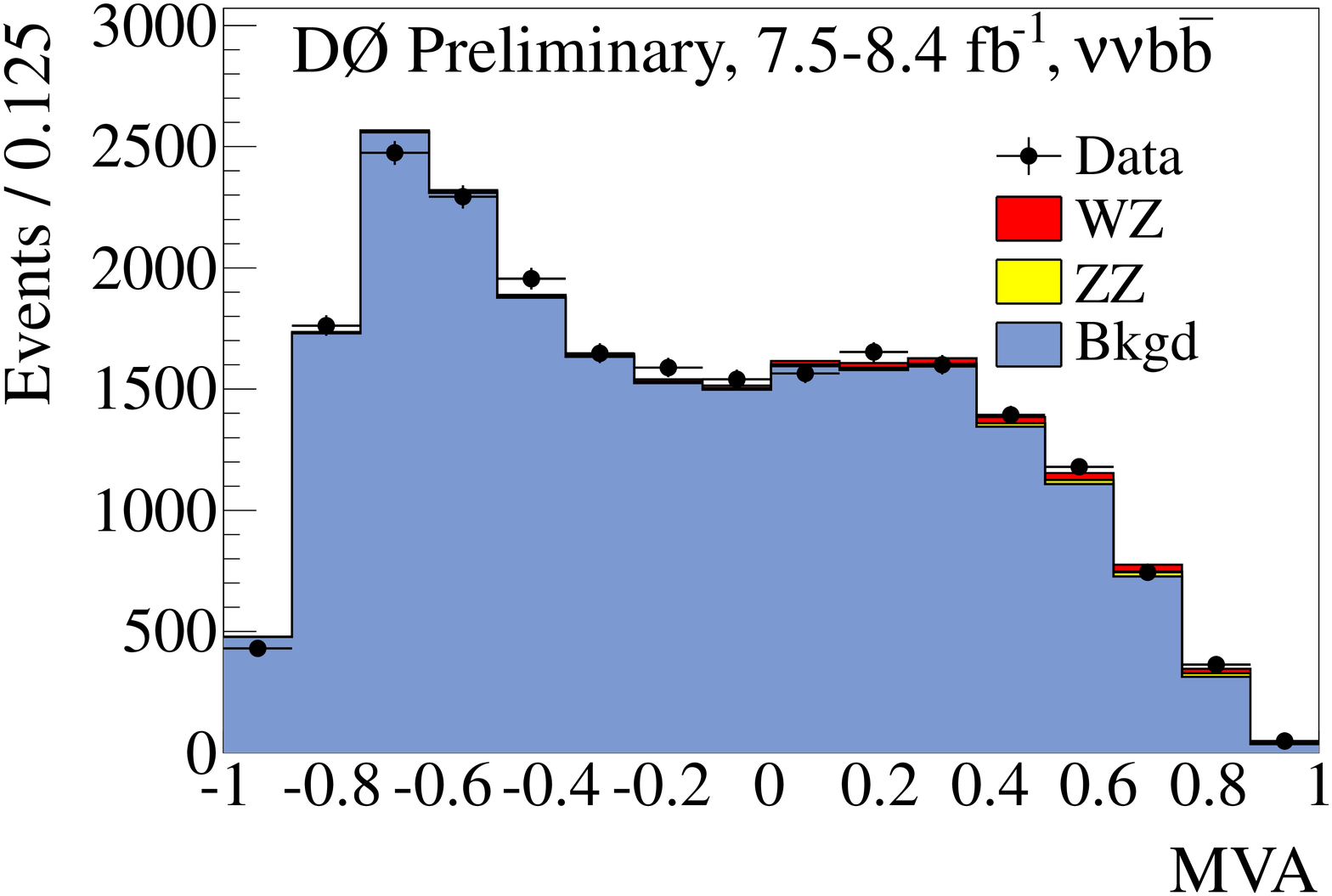} & 
\includegraphics[width=2.4in]{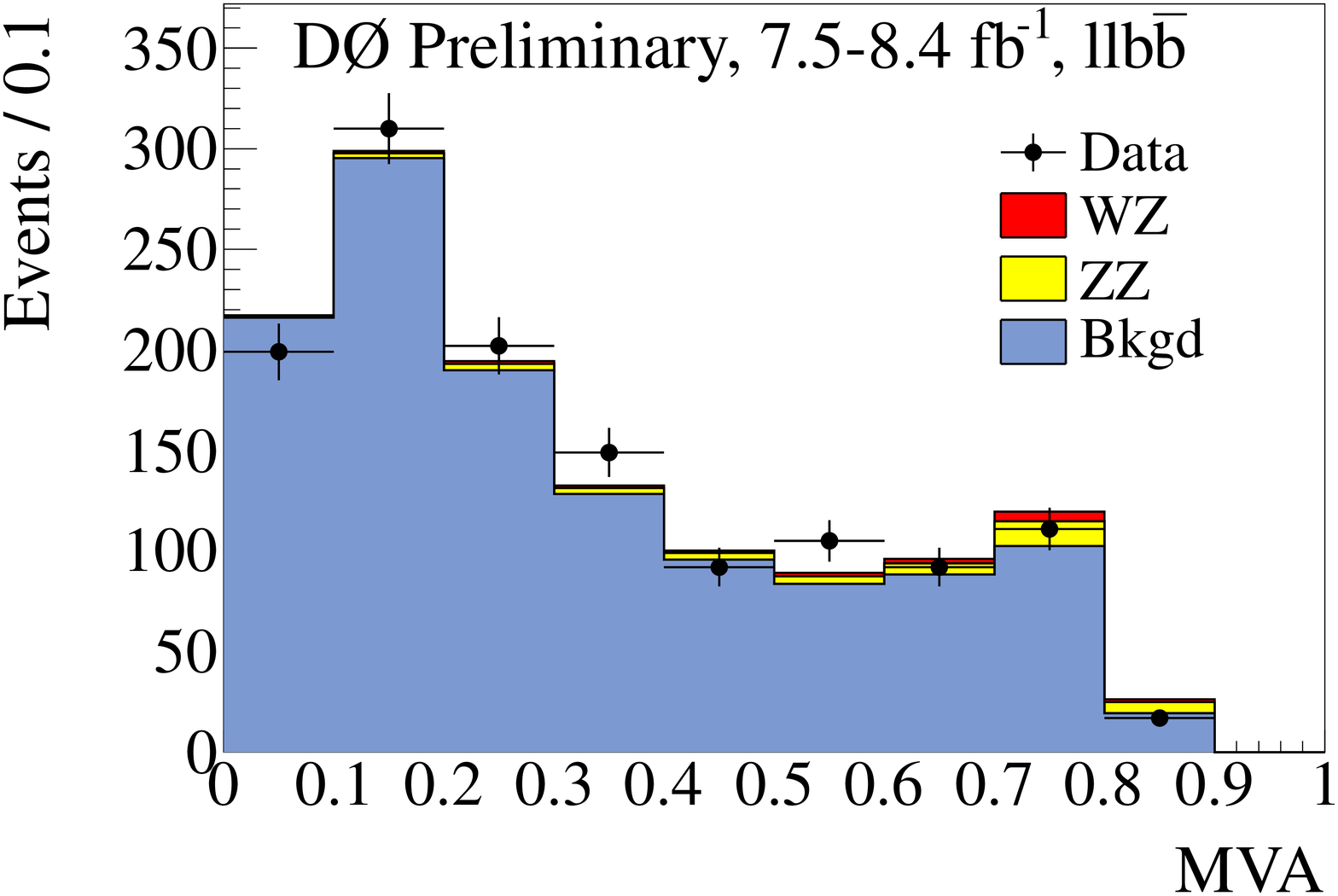} \\
{\bf (a)} & {\bf (b)}  & {\bf (c)} \\
\includegraphics[width=2.4in]{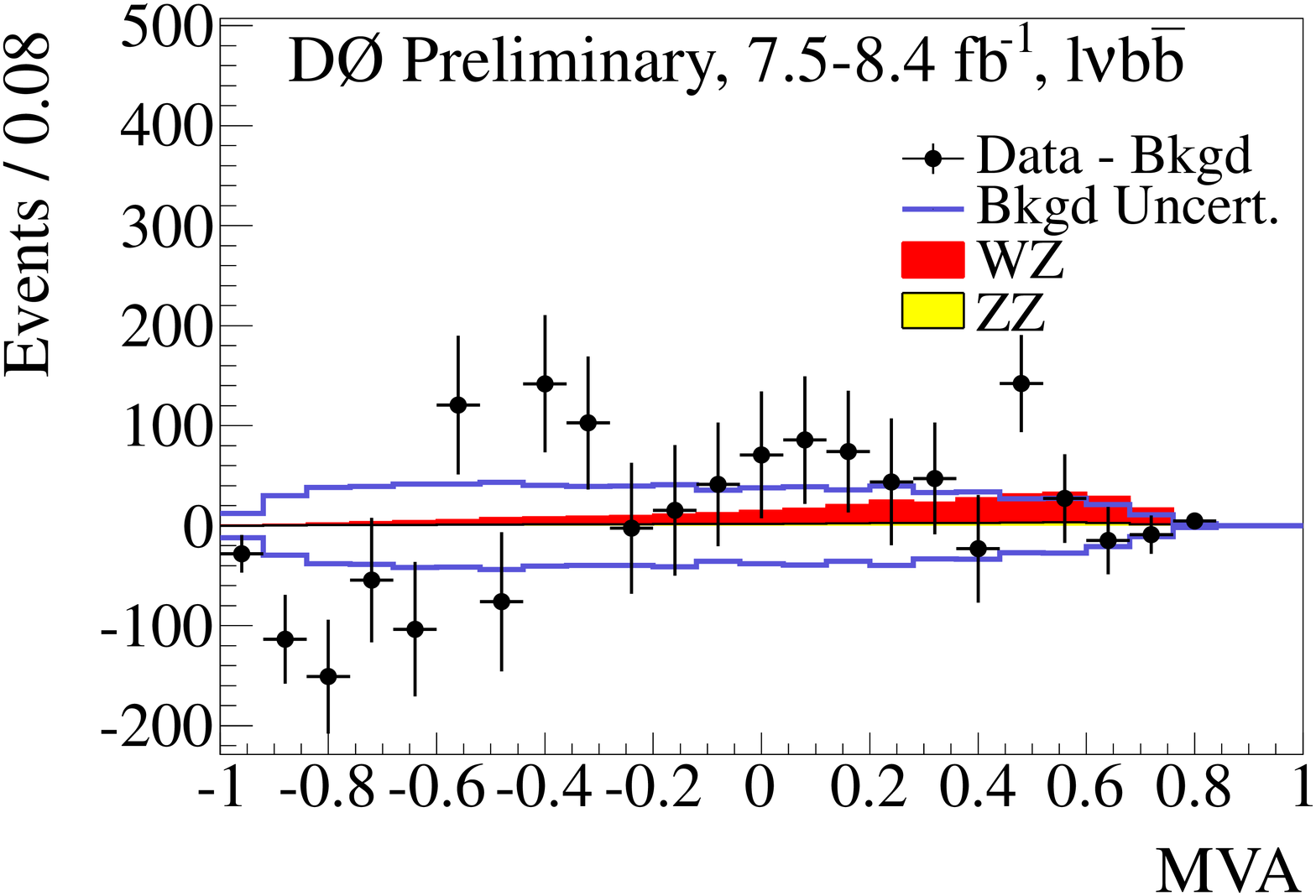}  &
\includegraphics[width=2.4in]{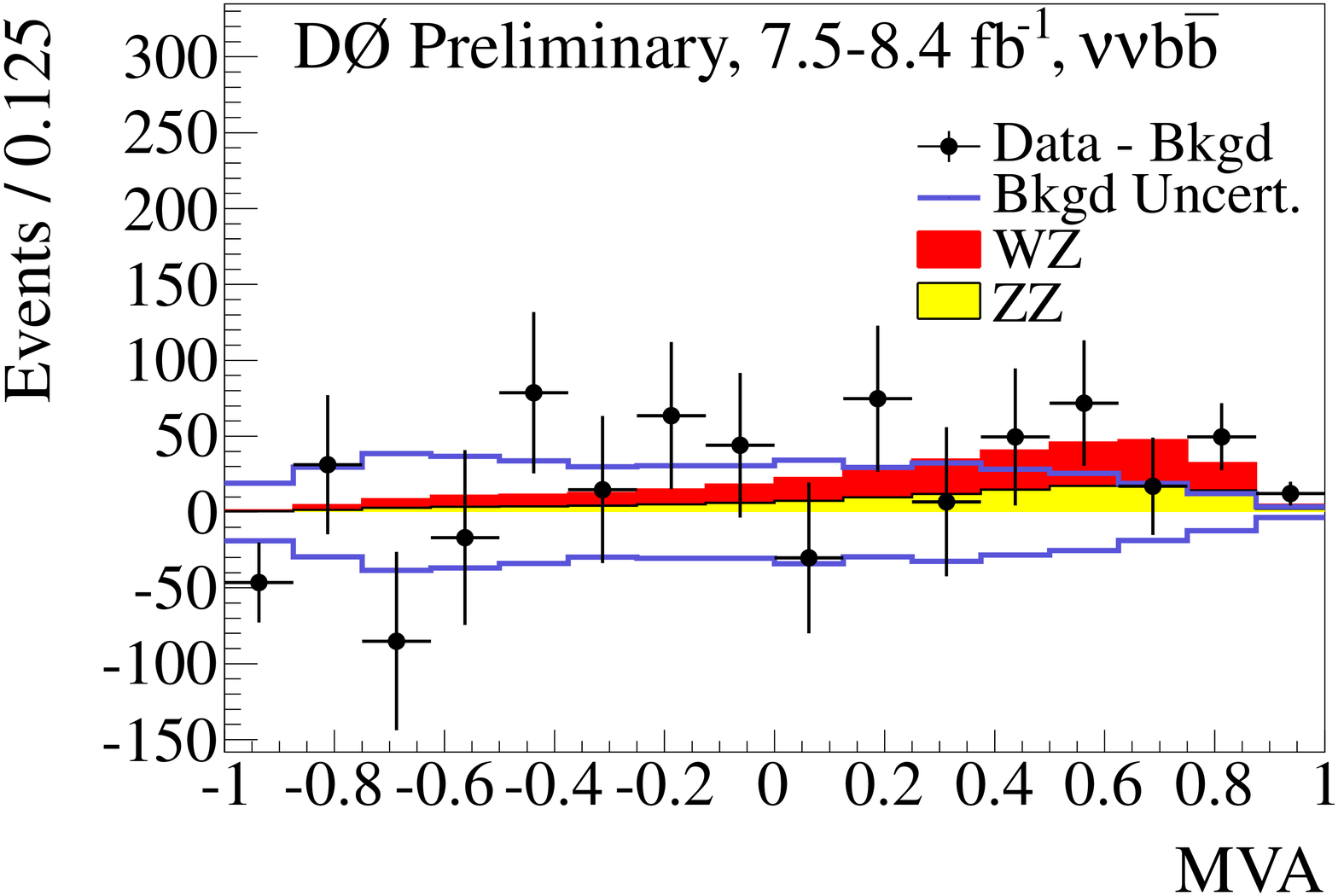} & 
\includegraphics[width=2.4in]{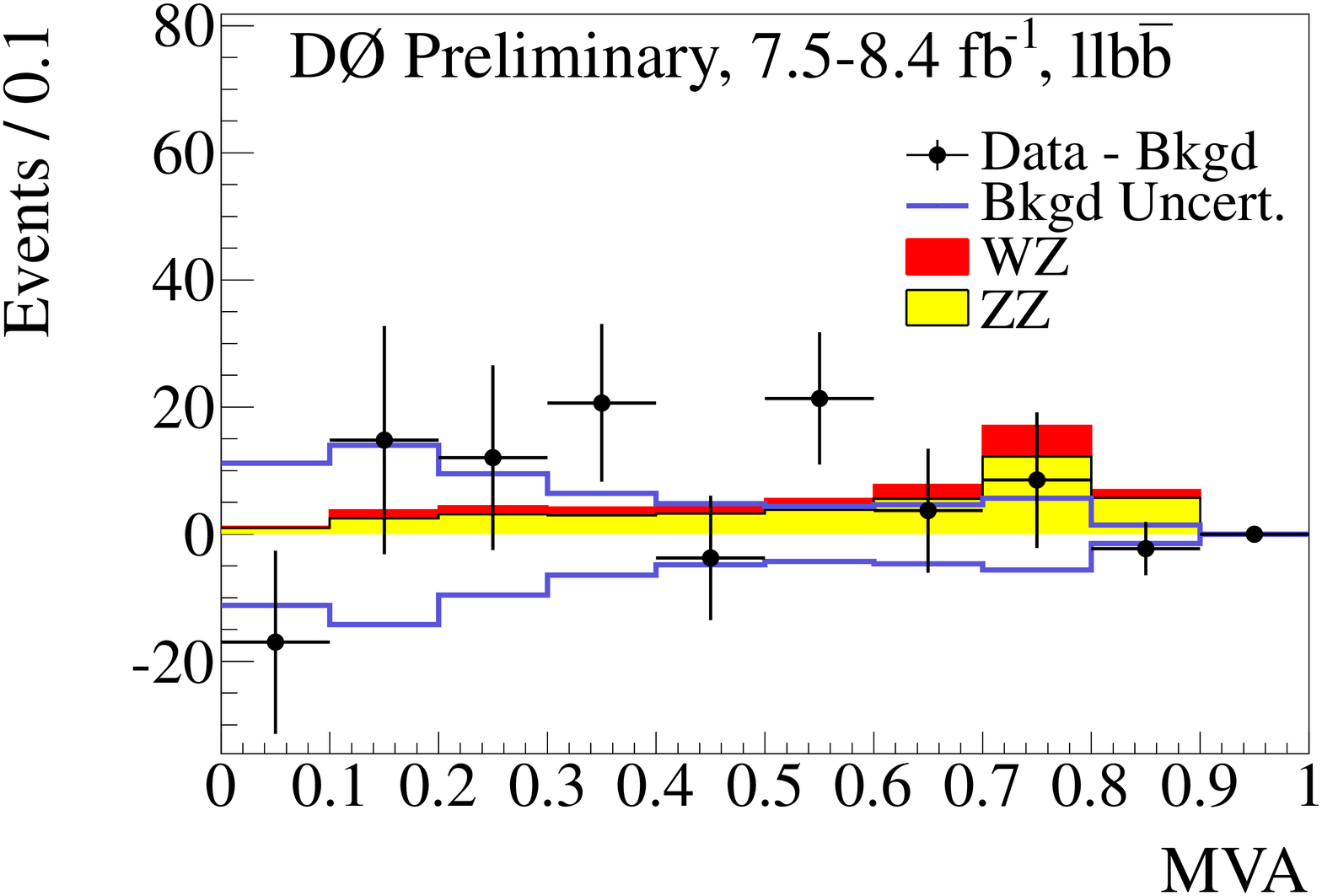} \\
{\bf (d)} & {\bf (e)}  & {\bf (f)}
\end{tabular} 
\end{centering} 
\caption{\label{fig:mva_chan} Comparison of fitted signal+background
  to the data in the final MVA distributions for the (a) \lvbb\ (b)
  \vvbb\ and (c) \llbb\ analyses (each summed over all sub-channels);
  and comparison of the measured signal to the background-subtracted
  data in the (d) \lvbb\ (e) \vvbb\ and (f) \llbb\ analyses.  The
  background has been fit to the data in the hypothesis that both
  signal and background are present.  Also shown is the $\pm$1
  standard deviation uncertainty on the fitted background.}
\end{figure*}

\begin{figure*}[tbp]
\begin{centering}
\begin{tabular}{ccc}
\includegraphics[width=2.4in]{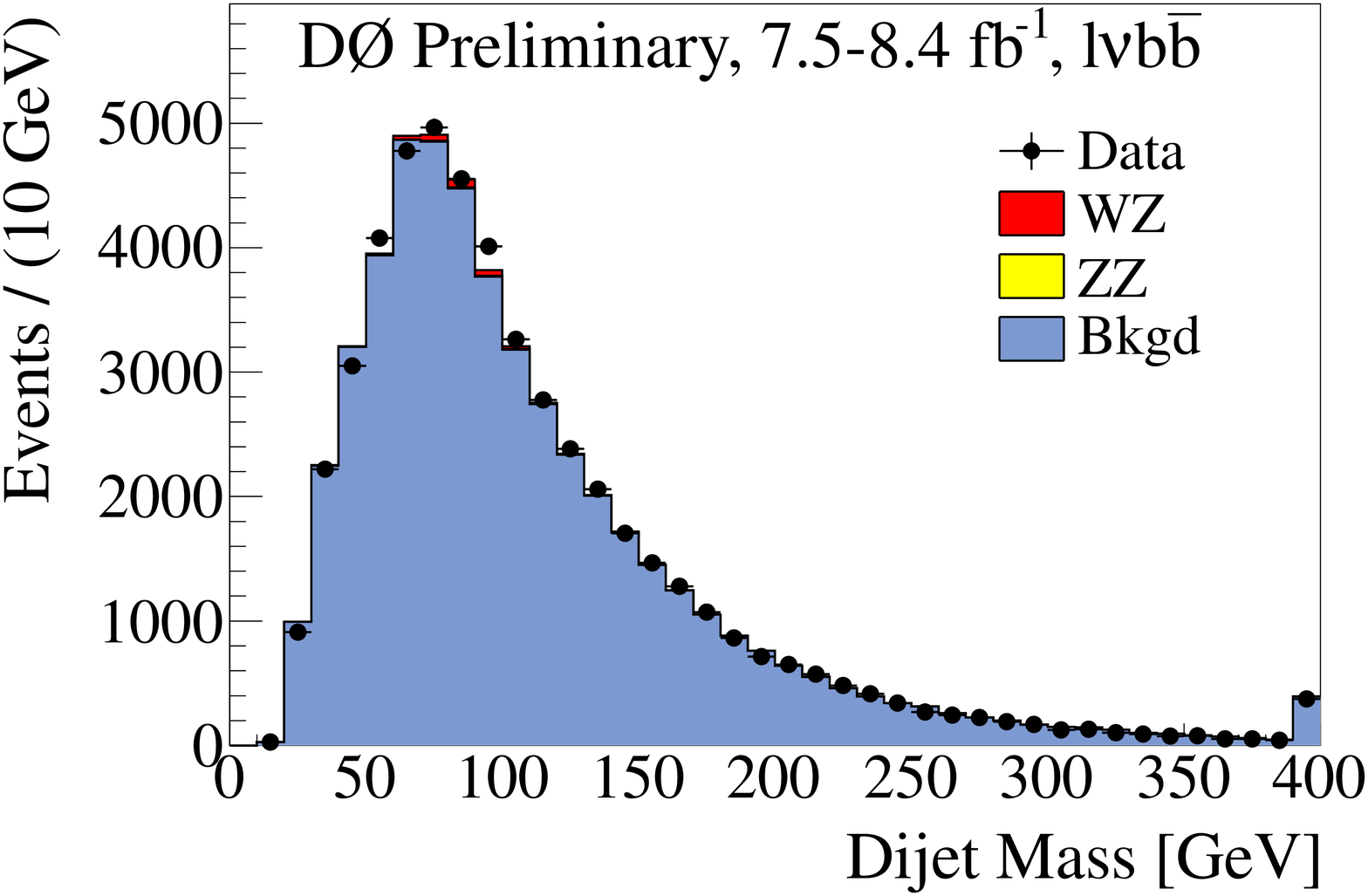}  &
\includegraphics[width=2.4in]{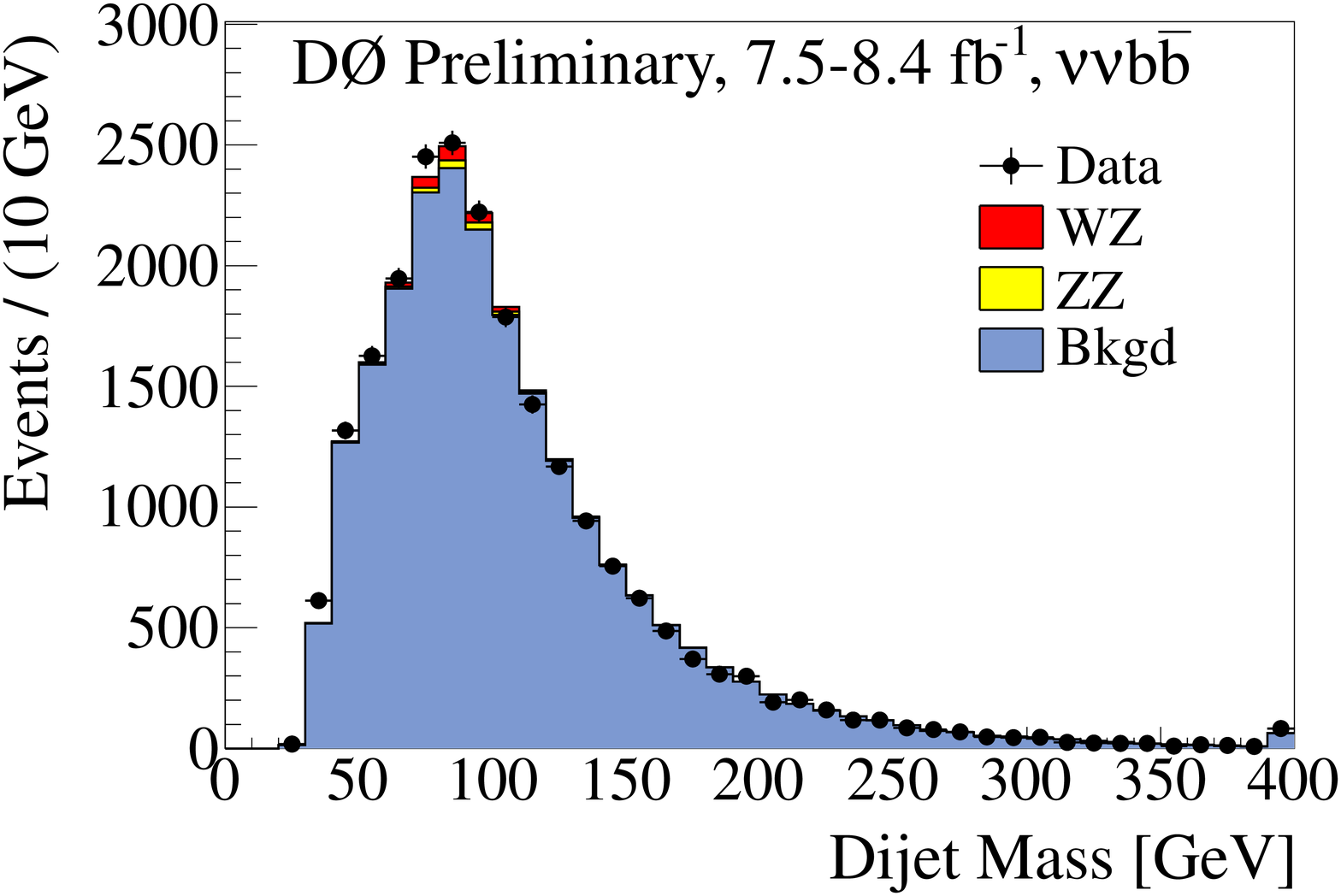} & 
\includegraphics[width=2.4in]{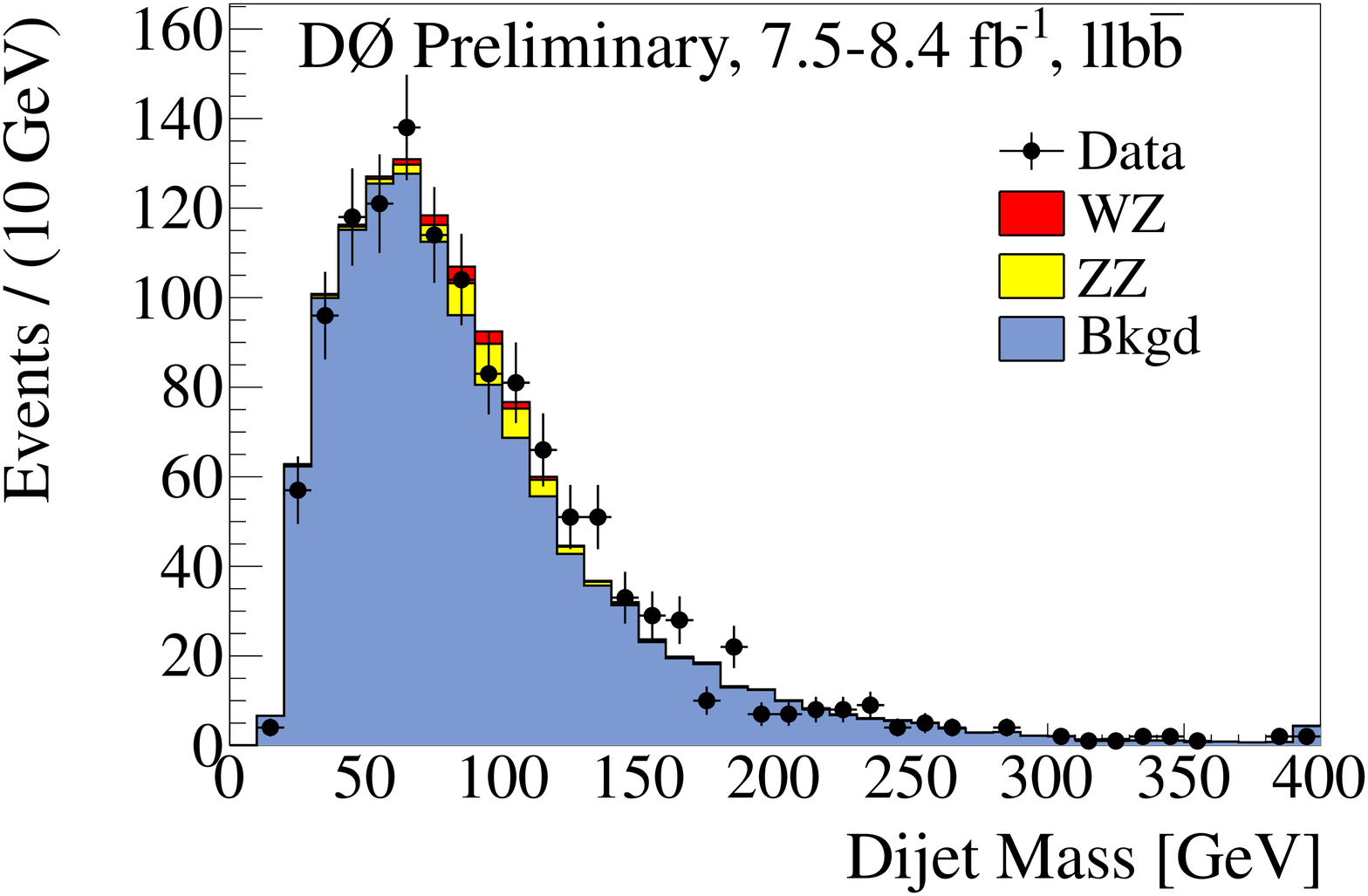} \\
{\bf (a)} & {\bf (b)}  & {\bf (c)}\\
\includegraphics[width=2.4in]{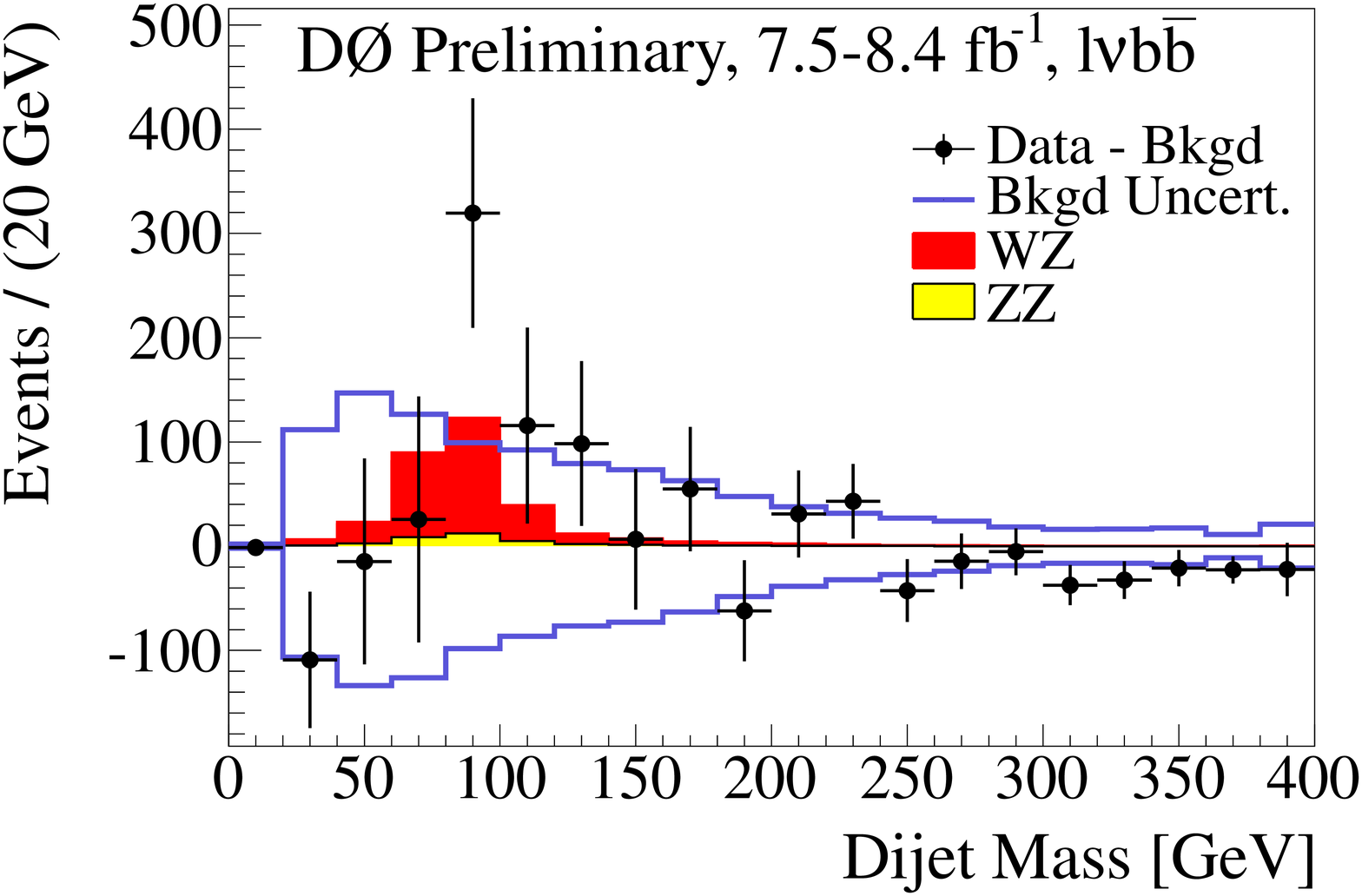}  &
\includegraphics[width=2.4in]{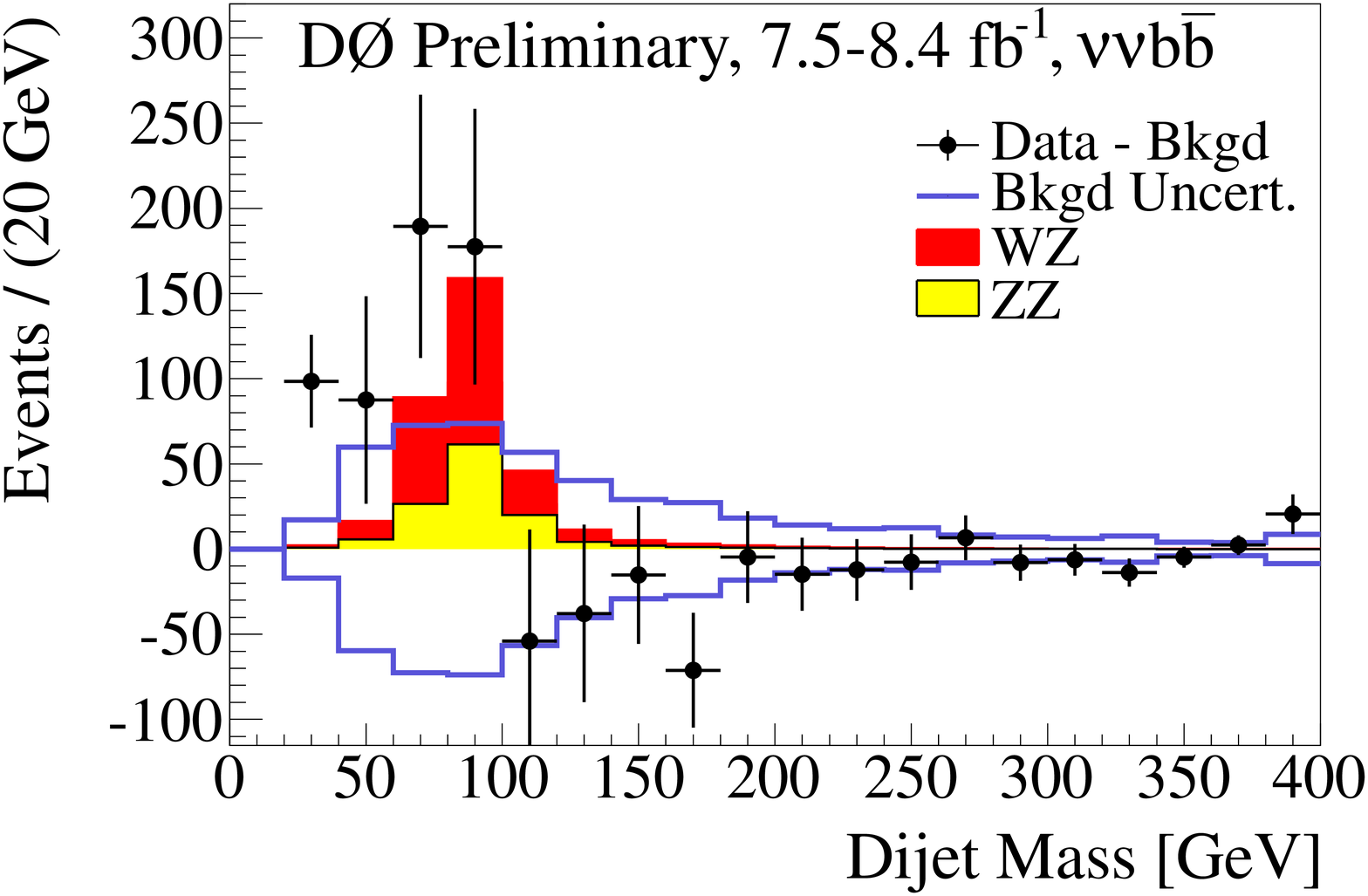} & 
\includegraphics[width=2.4in]{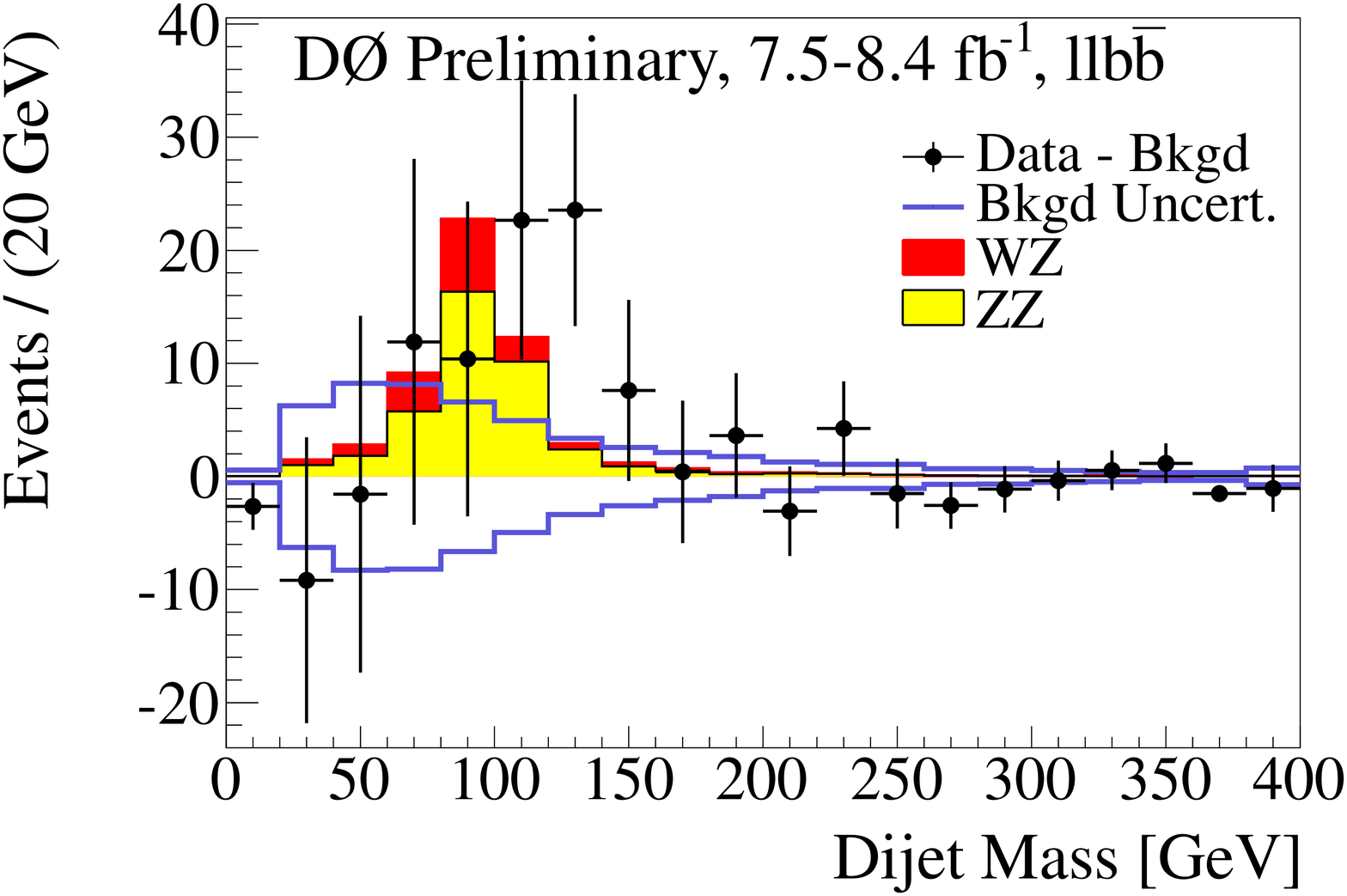} \\
{\bf (d)} & {\bf (e)}  & {\bf (f)}
\end{tabular} 
\end{centering} 
\caption{\label{fig:mjj_chan} Comparison of fitted signal+background
  to the data in the dijet mass distributions for the (a) \lvbb\ (b)
  \vvbb\ and (c) \llbb\ analyses (each summed over all sub-channels);
  and comparison of the measured signal to the background-subtracted
  data in the (d) \lvbb\ (e) \vvbb\ and (f) \llbb\ analyses.  The
  background has been fit to the data in the hypothesis that both
  signal and background are present.  Also shown is the $\pm$1
  standard deviation uncertainty on the fitted background.  Events
  with a dijet mass greater than 400 GeV are included in the last bin
  of the distribution.}
\end{figure*}

\end{document}

%% file: acknowledgement.tex
%
We thank the staffs at Fermilab and collaborating institutions,
and acknowledge support from the
DOE and NSF (USA);
CEA and CNRS/IN2P3 (France);
FASI, Rosatom and RFBR (Russia);
CNPq, FAPERJ, FAPESP and FUNDUNESP (Brazil);
DAE and DST (India);
Colciencias (Colombia);
CONACyT (Mexico);
KRF and KOSEF (Korea);
CONICET and UBACyT (Argentina);
FOM (The Netherlands);
STFC and the Royal Society (United Kingdom);
MSMT and GACR (Czech Republic);
CRC Program and NSERC (Canada);
BMBF and DFG (Germany);
SFI (Ireland);
The Swedish Research Council (Sweden);
and
CAS and CNSF (China).

%% file: lvbb-sys.tex

\begin{table}[h]
\begin{center}
  \caption{\label{tab:d0systwh} Systematic uncertainties for the
    $\ell\nu b{\bar{b}}$ single tag (ST) and double tag (DT) channels.
    Systematic uncertainties are listed by name; see the original
    references for a detailed explanation of their meaning and on how
    they are derived.  Uncertainties are relative, in percent on the
    event yield.  Shape uncertainties are labeled with an ``(S)'', and
    ``SO'' represents uncetrainties that affect only the shape, but
    not the event yield.  }
\vskip 0.2cm
{\centerline{$\ell\nu b\bar{b}$ Single Tag (ST) channels relative uncertainties (\%)}}
\vskip 0.099cm
\begin{ruledtabular}
\begin{tabular}{l c c c c c c }\\
Contribution                   &~Dibosons~ & $W+b\bar{b}/c\bar{c}$& $W$+l.f. & $~~~t\bar{t}~~~$ &single top&Multijet\\
\hline
Luminosity                     &  6.1    &  6.1  &    6.1  &    6.1  &    6.1  &   -- \\ 
Electron ID/Trigger efficiency   (S) & 1--5    & 2--4  &    2--4 &   1--2  &   1--2  &   -- \\       
Muon Trigger efficiency (S)          &  1--3   &  1--2 &    1--3 &    2--5 &    2--3 &   -- \\       
Muon ID efficiency/resolution       &   4.1   &   4.1 &     4.1 &     4.1 &     4.1 &   -- \\        
Jet ID efficiency  (S)          &  2--5   &  1--2 &    1--3 &    3--5 &    2--4 &   -- \\ 
Jet Energy Resolution (S)          &  4--7   &  1--3 &    1--4 &    2--5 &    2--4 &   -- \\       
Jet Energy Scale  (S)          &  4--7   &  2--5 &    2--5 &    2--5 &    2--4 &   -- \\       
Vertex Conf. Jet  (S)          & 4--10   & 5--12 &   4--10 &   7--10 &   5--10 &   -- \\       
$b$-tag/taggability (S)        & 1--4    &  1--2 &   3--7  &    3--5 &    1--2 &   -- \\ 
Heavy-Flavor K-factor          &   --    &    20 &      -- &   --    &   --    &   -- \\       
Multijet model, $e\nu b\bar{b}$ (S)   & 1--2    & 2--4  &    1--3 & 1--2    &    1--3 &   15 \\ 
Multijet model, $\mu\nu b\bar{b} $    &  --     &   2.4 &   2.4   &   --    &   --    &   20 \\ 
Cross Section                  &     6   &     9 &     9   &    10   &      10 &   -- \\ 
ALPGEN MLM pos/neg(S)          &   --    &   SO  &      -- &   --    &   --    &   -- \\       
ALPGEN Scale (S)               &   --    &   SO  &    SO   &   --    &   --    &   -- \\       
Underlying Event (S)           &   --    &   SO  &      -- &   --    &   --    &   -- \\       
PDF, reweighting               &  2      &  2    & 2       & 2       &    2    &   -- \\
\end{tabular}
\end{ruledtabular}
\vskip 0.5cm
{\centerline{$\ell\nu b\bar{b}$ Double Tag (DT) channels relative uncertainties (\%)}}
\vskip 0.099cm
\begin{ruledtabular}
\begin{tabular}{ l c c c c c c }   \\
Contribution  &~Dibosons~&$W+b\bar{b}/c\bar{c}$&$W$+l.f.&$~~~t\bar{t}~~~$&single top&Multijet \\
\hline
Luminosity                    &  6.1  &  6.1  &  6.1  &  6.1  &  6.1  &   --     \\ 
Electron ID/Trigger  efficiency (S)  & 2--5  & 2--3  &  2--3 & 1--2  & 1--2  &   --     \\       
Muon Trigger efficiency (S)         &  2--4 &  1--2 &  1--2 &  2--4 &  1--3 &   --     \\       
Muon ID efficiency/resolution      &   4.1 &   4.1 &   4.1 &   4.1 &   4.1 &   --     \\        
Jet ID efficiency  (S)         &  2--8 &  2--5 &  4--9 &  3--7 &  2--4 &   --     \\ 
Jet Energy Resolution    (S)         &  4--7 &  2--7 &  2--7 &  2--9 &  2--4 &   --     \\       
Jet Energy Scale  (S)         &  4--7 &  2--6 &  2--7 &  2--6 &  2--7 &   --     \\       
Vertex Conf. Jet  (S)         & 4--10 & 5--12 & 4--10 & 7--10 & 5--10 &   --     \\       
$b$-tag/taggability (S)       & 3--7  &  4--6 & 3--10 & 5--10 & 4--10 &   --     \\ 
Heavy-Flavor K-factor         &   --  &    20 &    -- &  --   &  --   &   --     \\       
Multijet model, $e\nu\bb$ (S)  & 1--2  & 2--4  & 1--3  & 1--2  &  1--3 &   15     \\ 
Multijet model, $\mu\nu\bb$   &  --   &   2.4 &   2.4 & --    &  --   &   20     \\ 
Cross Section                 &     6 &     9 &     9 &    10 &    10 &   --     \\
ALPGEN MLM pos/neg(S)         &   --  &   SO  &    -- &   --  &   --  &   --     \\       
ALPGEN Scale (S)              &   --  &   SO  &    SO &   --  &   --  &   --     \\       
Underlying Event (S)          &   --  &   SO  &    -- &   --  &   --  &   --     \\       
PDF, reweighting              &  2    &  2    & 2     & 2     &  2    &   --     \\
\end{tabular}
\end{ruledtabular}

\end{center}
\end{table}

%% file: vvbb-sys.tex
\begin{table}[h]
  \caption{\label{tab:d0systzhll} Systematic uncertainties for the $\nu\nu b{\bar{b}}$
    single tag (ST) and double tag (DT) channels. Systematic uncertainties
    are listed by name; see the original references for a detailed
    explanation of their meaning and on how they are derived. Uncertainties
    are relative, in percent on the event yield.  Shape uncertainties are
    labeled with an ``(S)'', and ``SO'' represents shape only uncertainty.}
\vskip 0.2cm
{\centerline{$\nu\nu b\bar{b}$ Single Tag (ST) channels relative uncertainties (\%)}}
\vskip 0.099cm
\begin{ruledtabular}
\begin{tabular}{l c c c c  c }\\
Contribution            & Top  & $V+b\bar{b}/c\bar{c}$ & $V$+l.f. & Dibosons & Multijet \\
\hline
Jet ID efficiency (S)      & 2.0  &  2.0   &  2.0   &  2.0  & -- \\
Jet Energy Scale (S)            & 2.2  &  1.6   &  3.1   &  1.0  & -- \\
Jet Energy Resolution (S)              & 0.5  &  0.3   &  0.3   &  0.9  & -- \\
Vertex Conf. / Taggability (S)  & 3.2  &  1.9   &  1.7   &  1.8  & -- \\
b Tagging (S)                   & 1.1  &  0.8   &  1.8   &  1.2  & -- \\
Lepton Identification           & 1.6  &  0.9   &  0.8   &  1.0  & -- \\
Trigger                         & 2.0  &  2.0   &  2.0   &  2.0  & -- \\
Heavy Flavor Fractions          & --   &  20.0  &  --    &  --   & -- \\
Multijet model                  & --   &  --    &  --    &  --   & 25 \\
Cross Sections                  & 10.0 &  10.2  &  10.2  &  7.0  & -- \\
Luminosity                      & 6.1  &  6.1   &  6.1   &  6.1  & -- \\
Multijet Normalilzation         & --   &  --    &  --    &  --   & -- \\
ALPGEN MLM (S)                  & --   &  --    &  SO    &  --   & -- \\
ALPGEN Scale (S)                & --   &  SO    &  SO    &  --   & -- \\
Underlying Event (S)            & --   &  SO    &  SO    &  --   & -- \\
PDF, reweighting (S)            & SO   &  SO    &  SO    &  SO   & -- \\
\end{tabular}
\vskip 0.5cm
{\centerline{$\nu\nu b\bar{b}$ Double Tag (DT) channels relative uncertainties (\%)}}
\vskip 0.099cm
\begin{tabular}{ l c c c c c }   \\
Contribution            & Top  & $V+b\bar{b}/c\bar{c}$ & $V$+l.f. & Dibosons & Multijet \\
\hline
Jet ID efficiency          & 2.0  &  2.0   &  2.0   &  2.0   & -- \\
Jet Energy Scale                & 2.1  &  1.6   &  3.4   &  1.2   & -- \\
Jet Energy Resolution                  & 0.7  &  0.4   &  0.5   &  1.5   & -- \\
Vertex Conf. / Taggability      & 2.6  &  1.6   &  1.6   &  1.8   & -- \\
b Tagging                       & 6.2  &  4.3   &  4.3   &  3.7   & -- \\
Lepton Identification           & 2.0  &  0.9   &  0.8   &  0.9   & -- \\
Trigger                         & 2.0  &  2.0   &  2.0   &  2.0   & -- \\
Heavy Flavor Fractions          & --   &  20.0  &  --    &  --    & -- \\
Multijet model                  & --   &  --    &  --    &  --   & 25 \\
Cross Sections                  & 10.0 &  10.2  &  10.2  &  7.0   & -- \\
Luminosity                      & 6.1  &  6.1   &  6.1   &  6.1   & -- \\
Multijet Normalilzation         & --   &  --    &  --    &  --    & -- \\
ALPGEN MLM pos/neg (S)          &  --  &  --    &  SO    &  --    & -- \\
ALPGEN Scale (S)                &  --  &  SO    &  SO    &  --    & -- \\
Underlying Event (S)            &  --  &  SO    &  SO    &  --    & -- \\
PDF, reweighting (S)            &  SO  &  SO    &  SO    &  SO    & -- \\

\end{tabular}
\end{ruledtabular}

\end{table}

%% file: llbb-sys.tex
\begin{table}
  \caption{\label{tab:d0llbb1} Systematic uncertainties for the \llbb\ 
    single tag (ST) and double tag (DT) channels. Systematic uncertainties are listed by name; see the original
    references for a detailed explanation of their meaning and on how they are derived.
    Uncertainties are relative, in percent on the event yield. Shape uncertainties are
    labeled with an ``(S)''. }
\vskip 0.2cm
{\centerline{$\ell\ell b \bar{b}$ Single Tag (ST) channels relative uncertainties (\%)}}
\vskip 0.099cm
\begin{ruledtabular}
\begin{tabular}{  l  c  c  c  c  c  c  c }   
Contribution               & Multijet& $Z$+l.f.  &  $Z+\bb$ & $Z+\cc$ & Dibosons & Top\\ \hline
Jet Energy Scale (S)       &   --    &  3.0   &  8.4   &  10   &  3.3   &  1.5  \\
Jet Energy Resolution (S)  &   --    &  3.9   &  5.2   &  5.3  & 0.04   &  0.6  \\
Jet ID efficiency (S)      &   --    &  0.9   &  0.6   &  0.2  &  1.0   &  0.3  \\
Taggability (S)            &   --    &  5.2   &  7.2   &  7.3  &  6.9   &  6.5  \\
$Z p_T$ Model (S)          &   --    &  2.7   &  1.4   &  1.5  &   --   &   --  \\
HF Tagging Efficiency (S)  &   --    &   --   &  5.0   &  9.4  &   --   &  5.2  \\
LF Tagging Efficiency (S)  &   --    &   73   &   --   &   --  &  5.8   &   --  \\
$ee$ Multijet Shape (S)    &   53    &   --   &   --   &   --  &   --   &   --  \\
Multijet Normalization     &  20-50  &   --   &   --   &   --  &   --   &   --  \\
$Z$+jets Jet Angles (S)    &   --    &  1.7   &  2.7   &  2.8  &   --   &   --  \\
Alpgen MLM (S)             &   --    &  0.3   &   --   &   --  &   --   &   --  \\
Alpgen Scale (S)           &   --    &  0.4   &  0.2   &  0.2  &   --   &   --  \\
Underlying Event (S)       &   --    &  0.2   &  0.1   &  0.1  &   --   &   --  \\
Trigger (S)                &   --    &  0.03  &  0.2   &  0.3  &  0.3   &  0.4  \\
Cross Sections             &   --    &   --   &  20    &  20   &  7     &  10   \\
Normalization              &   --    &  1.3   &  1.3   &  1.3  &  8.0   &  8.0  \\
PDFs                       &   --    &  1.0   &  2.4   &  1.1  &  0.7   &  5.9 
\end{tabular}
\end{ruledtabular}
\vskip 0.5cm

{\centerline{$\ell\ell b \bar{b}$ Double Tag (DT) channels relative uncertainties (\%)}}
\vskip 0.099cm
\begin{ruledtabular}
\begin{tabular}{  l  c  c  c  c  c  c  c }  \\
Contribution               & Multijet& $Z$+l.f.  &  $Z+\bb$ & $Z+\cc$ & Dibosons & Top\\  \hline
Jet Energy Scale (S)       &   --    &  4.0   &  6.4   &  8.2   &  3.8   &  2.7  \\
Jet Energy Resolution(S)   &   --    &  2.6   &  3.9   &  4.1   &  0.9   &  1.5  \\
JET ID efficiency (S)      &   --    &  0.7   &  0.3   &  0.2   &  0.7   &  0.4  \\
Taggability (S)            &   --    &  8.6   &  6.5   &  8.2   &  4.6   &  2.1  \\
$Z_{p_T}$ Model (S)        &   --    &  1.6   &  1.3   &  1.4   &   --   &   --  \\
HF Tagging Efficiency (S)  &   --    &   --   &  1.3   &  3.2   &   --   &  0.7  \\
LF Tagging Efficiency (S)  &   --    &   72   &   --   &   --   &  4.0   &   --  \\
$ee$ Multijet Shape (S)    &    59   &   --   &   --   &   --   &   --   &   --  \\
Multijet Normalization     &  20-50  &   --   &   --   &   --   &   --   &   --  \\
$Z$+jets Jet Angles (S)    &   --    &  2.0   &  1.5   &  1.5   &   --   &   --  \\
Alpgen MLM (S)             &   --    &  0.4   &   --   &   --   &   --   &   --  \\
Alpgen Scale (S)           &   --    &  0.2   &  0.2   &  0.2   &   --   &   --  \\
Underlying Event(S)        &   --    &  0.1   & 0.02   &  0.1   &   --   &   --  \\
Trigger (S)                &   --    &  0.3   &  0.2   &  0.1   &  0.2   &  0.5  \\
Cross Sections             &   --    &   --   & 20     & 20     & 7      & 10    \\
Normalization              &   --    &  1.3   & 1.3    & 1.3    & 8.0    & 8.0   \\
PDFs                       &   --    &  1.0   & 2.4    & 1.1    & 0.7    & 5.9 
\end{tabular}
\end{ruledtabular}
\end{table}